\documentclass[lettersize,journal]{IEEEtran}
\usepackage{amsmath,amssymb,amsfonts}
\usepackage{algorithm}
\usepackage{algpseudocode}
\usepackage{array}
\usepackage[caption=false,font=normalsize,labelfont=sf,textfont=sf]{subfig}
\usepackage{textcomp}
\usepackage{stfloats}
\usepackage{url}
\usepackage{verbatim}
\usepackage{graphicx}
\usepackage{cite}
\usepackage{booktabs}
\usepackage{xcolor}
\usepackage{epsfig}
\usepackage{epstopdf}
\usepackage{acronym}
\graphicspath{{fig/}}
\usepackage{mathtools}
\usepackage{amsthm}
\usepackage{bm}

\usepackage{ragged2e}

\newacro{ACDD}{Alamouti with cyclic delay diversity}
\newacro{URLLC}{ultra-reliable low-latency communications}
\newacro{3GPP}{third generation partnership project}
\newacro{PHY}{physical layer}
\newacro{MIMO}{multiple-input multiple-output}
\newacro{SIMO}{single-input multiple-output}
\newacro{MISO}{multiple-input single-output}
\newacro{SISO}{single-input single-output}
\newacro{MRC}{maximum-ratio combining}
\newacro{SNR}{signal-to-noise ratio}
\newacro{CP}{cyclic prefix}
\newacro{CDD}{cyclic delay diversity}
\newacro{FSC}{frequency-selective channel}
\newacro{STC}{space-time coding}
\newacro{FFT}{fast Fourier transform}
\newacro{LMMSE}{linear minimum mean-squared error}
\newacro{FER}{frame error rate}
\newacro{OFDM}{orthogonal frequency division multiplexing}
\newacro{OCDM}{orthogonal chirp division multiplexing}
\newacro{FSC}{frequency-selective channel}
\newacro{CSI}{channel state information}
\newacro{LMMSE-PIC}{linear minimum mean squared error with parallel interference cancellation}
\newacro{PFE}{perfect-feedback equalizer}
\newacro{FD}{full-duplex}
\newacro{PDP}{power delay profile}
\newacro{PDF}{probability density function}
\newacro{DFT}{discrete Fourier transform}
\newacro{SDFT}{sparse DFT}
\newacro{ICI}{inter-carrier interference}
\newacro{OTFS}{orthogonal time frequency space}
\newacro{AWGN}{additive white Gaussian noise}
\newacro{SWH}{sparse Walsh-Hadamard}
\newacro{LLR}{log-likelihood ratio}
\newacro{PMF}{probability mass function}
\newacro{CRC}{cyclic redundancy check}
\newacro{PAM}{pulse amplitude modulation}
\newacro{QAM}{quadrature amplitude modulation}
\newacro{FWHT}{fast Walsh-Hadamard transform}
\newacro{MAP}{maximum a-posteriori}
\newacro{SC}{single-carrier}
\newacro{ISI}{inter-symbol interference}
\newacro{ZP}{zero-padding}
\newacro{BCJR}{Bahl, Cocke, Jelinek, and Raviv}
\newacro{WHT}{Walsh-Hadamard transform}
\newacro{APP}{a-posteriori probability}
\newacro{SILE-EPIC}{self-iterated linear equalizer with expectation propagation}
\newacro{EP}{expectation propagation}
\newacro{i.i.d.}{independent and identically distributed}
\newacro{CWCU}{component wise conditionally unbiased}
\newacro{MSE}{mean squared error}
\newacro{EXIT}{extrinsic information transfer}
\newacro{MI}{mutual information}
\newacro{PAPR}{peak-to-average power ratio}
\newacro{DFT-s}{discrete Fourier transform-spread}
\newacro{AMP}{approximate message passing}
\newacro{GAMP}{generalized \ac{AMP}}
\newacro{VAMP}{vector \ac{AMP}}
\newacro{RSC}{recursive systematic convolutional}
\newacro{QPSK}{quadrature phase-shift keying}
\newacro{CFAR}{constant false alarm rate}
\newacro{PD}{probability of detection}
\newacro{PFA}{probability of false alarm}
\newacro{RV}{random variable}
\newacro{CDF}{cumulative distribution function}
\newacro{HD-ZP}{half-duplex ZP}
\newacro{FD-CP}{full-duplex ZP}
\newacro{DFRC}{dual-function radar communication}
\newacro{SINR}{signal-to-interference noise ratio}
\newacro{ISAC}{integrated sensing and communication}
\newacro{SI}{self-interference}
\newacro{RSI}{residual self-interference}
\newacro{ADC}{analog-to-digital converter}
\newacro{ED}{energy-detection}
\newacro{IDFT}{inverse discrete Fourier Transform}
\newacro{SFFT}{symplectic finite Fourier transform }
\newacro{CRB}{Cram{\'{e}}r-Rao bound}
\newacro{ZC}{Zadoff-Chu}
\newacro{RMSE}{root mean square error}
\newacro{UW}{unique word}
\newacro{GFDM}{generalized frequency division multiplexing}
\newacro{RRC}{root-raised cosine}
\newacro{UB}{upper bound}
\newacro{CEF}{channel estimation field}
\newacro{CFO}{carrier frequency offset}
\newacro{CPI}{coherent processing interval}
\newacro{ULA}{uniform linear array}
\newacro{TO}{time-offset}

\def\BibTeX{{\rm B\kern-.05em{\sc i\kern-.025em b}\kern-.08em
		T\kern-.1667em\lower.7ex\hbox{E}\kern-.125emX}}
	
\usepackage{soul,color}

\usepackage{tikz}
\usepackage{pgfplots}
\usetikzlibrary{shapes,arrows}
\usetikzlibrary{positioning,calc}
\usetikzlibrary{decorations.pathreplacing,calligraphy}
\usetikzlibrary{arrows.meta}
\usepackage{siunitx}
\usetikzlibrary{shadings,patterns}

\tikzset{add/.style n args={4}{
		minimum width=3mm,
		path picture={
			\draw[black] 
			(path picture bounding box.south east) -- (path picture bounding box.north west)
			(path picture bounding box.south west) -- (path picture bounding box.north east);
			\node at ($(path picture bounding box.south)+(0,0.13)$)     {\tiny #1};
			\node at ($(path picture bounding box.west)+(0.13,0)$)      {\tiny #2};
			\node at ($(path picture bounding box.north)+(0,-0.13)$)        {\tiny #3};
			\node at ($(path picture bounding box.east)+(-0.13,0)$)     {\tiny #4};
		}
	}
}

\tikzset{add2/.style n args={4}{
		minimum width=1mm,
		path picture={
			\draw[black] 
			(path picture bounding box.south) -- (path picture bounding box.north)
			(path picture bounding box.west) -- (path picture bounding box.east);
			\node at ($(path picture bounding box.south)+(0,0.13)$)     {\tiny #1};
			\node at ($(path picture bounding box.west)+(0.13,0)$)      {\tiny #2};
			\node at ($(path picture bounding box.north)+(0,-0.13)$)        {\tiny #3};
			\node at ($(path picture bounding box.east)+(-0.13,0)$)     {\tiny #4};
		}
	}
}

\usepackage{pgfplots}
\usepgfplotslibrary{polar}
\usepackage{tikz}

\newcounter{mytempeqncnt}

\definecolor{applegreen}{rgb}{0.55, 0.71, 0.0}
\definecolor{awesome}{rgb}{1.0, 0.08, 0.15}
\definecolor{azure(colorwheel)}{rgb}{0.0, 0.5, 1.0}
\definecolor{darklavender}{rgb}{0.45, 0.31, 0.59}
\definecolor{cyan(process)}{rgb}{0.0, 0.72, 0.92}
\definecolor{brightmaroon}{rgb}{0.76, 0.13, 0.28}
\definecolor{ao(english)}{rgb}{0.0, 0.5, 0.0}
\definecolor{brightturquoise}{rgb}{0.03, 0.91, 0.87}
\definecolor{bondiblue}{rgb}{0.0, 0.58, 0.71}
\definecolor{atomictangerine}{rgb}{1.0, 0.6, 0.4}
\definecolor{classicrose}{rgb}{0.98, 0.8, 0.91}
\definecolor{copperrose}{rgb}{0.6, 0.4, 0.4}

\definecolor{lightgreen}{rgb}{0.56, 0.93, 0.56}
\definecolor{mediumgreen}{rgb}{0.0, 0.5, 0.0}
\definecolor{darkgreen}{rgb}{0.0, 0.35, 0.25}

\definecolor{lightpurple}{RGB}{229,204,255}
\definecolor{mediumpurple}{RGB}{147,112,219}
\definecolor{darkpurple}{RGB}{75,0,130}

\definecolor{deepskyblue}{rgb}{0, 0.75, 1.0}
\definecolor{dodgerblue}{rgb}{0.12, 0.56, 1.0}

\definecolor{wine}{rgb}{0.45, 0.18, 0.22}
\definecolor{crimson}{rgb}{0.86, 0.08, 0.24}
\definecolor{salmon}{rgb}{0.98, 0.5, 0.45}

\definecolor{goldenrod}{rgb}{0.85, 0.65, 0.13}
\definecolor{lightyellow}{rgb}{1.0, 1.0, 0.88}

\definecolor{beige}{rgb}{0.96, 0.96, 0.86}

\definecolor{lightblue}{RGB}{173, 216, 230}

\newcommand{\rev}[1]{\textcolor{black}{#1}}
\newcommand{\revv}[1]{\textcolor{black}{#1}}
\definecolor{my_color}{RGB}{0, 0, 255} 

\begin{document}

\title{Unique Word-Based Frame Design for Bistatic Integrated Sensing and Communication
 \thanks{Roberto Bomfin is with the Engineering Division, New York University
(NYU) Abu Dhabi, 129188, UAE (email: roberto.bomfin@nyu.edu).}
\thanks{Marwa Chafii is with Engineering Division, New York University (NYU)
Abu Dhabi, 129188, UAE and NYU WIRELESS, NYU Tandon School of
Engineering, Brooklyn, 11201, NY, USA (email: marwa.chafii@nyu.edu).}}

\author{Roberto Bomfin and Marwa Chafii,~\IEEEmembership{Senior Member,~IEEE}}



\maketitle
	
\begin{abstract}
Integrated sensing and communication (ISAC) aims at enhancing the network functionalities and enabling new applications in the upcoming communications networks. In this paper, we propose two unique word (UW)-based frame designs for bistatic ISAC. The approach consists of replacing the cyclic prefix (CP) with a Zadoff-Chu (ZC)-based sequence. With this approach, the radar receiver does not need to know the data symbols to perform sensing and the data rate is not compromised by the addition of extra pilots.
The sensing performance of the UW-based frames is compared with that of orthogonal frequency division multiplexing (OFDM) as well as the pilot-symbol (PS) based radar processing.
We derive the Cram{\'{e}}r-Rao bound (CRB) considering a band-limited system with raised-cosine filtering.
Furthermore, we provide low-complexity fast Fourier transform (FFT)-based radar receivers that perform integer and \rev{fine grid multi-target} delay-Doppler (DD) estimations.
For the integer FFT-based receiver, an upper bound for the outlier probability is derived when the true DD falls outside the integer grid.
The results demonstrate that the UW frames exhibit competitive radar performance with PS while having a 16.67\% higher data rate for the cases investigated.
\end{abstract}


\section{Introduction}
\IEEEPARstart{T}{he} recent advancements in wireless communications and radar technologies have been taking similar paths by employing antenna array techniques and moving toward higher frequency bands, leading to similarities in hardware architecture, channel characteristics, and signal processing.
This trend yields a unique opportunity for the integrated sensing and communication (ISAC) into the same network infrastructure, which has been considered as a new technology for future networks by academia and industry \cite{chafii2023twelve,LiuJSAC,ZhangISAC,BomfinSPAWC}.
Many applications that take advantage of ISAC have been considered, including smart factoring, the Internet of Things, robotics, environmental monitoring, and vehicular communications.


One of the major challenges of ISAC is the design of unified waveforms, whose objective is to use the same signal to perform both communication and sensing tasks.
In this regard, three categories have been conceived, namely, sensing-centric design (SCD), communication-centric design (CCD), and joint design (JD).
In short, the SCD approach incorporates communication on top of an existing sensing technology \cite{Shlezinger,Zhang_ISAC}.
Conversely, the CCD's goal is to include sensing in a communications-only system \cite{Sturm,Barneto,Johnston,Sayed2}.
Lastly, the JD's idea is to develop new waveforms along with beamforming techniques \cite{LiuTSP,Bazzi,BAZZI_papr}.
Each strategy has its pros and cons, and their adoption is dependent on the particular application and ISAC scenario. 
Additionally, there are many possibilities for the sensing architecture in ISAC, namely, uplink and downlink signaling, monostatic, bistatic, and multistatic sensing configurations. 
The choice of each configuration depends on the application, and multiple combinations of different configurations are possible.
For instance, the downlink-monostatic ISAC represents a typical ISAC scenario where the base station senses objects within its adjacency, which can be improved by uplink-bistatic sensing as demonstrated in \cite{Alhil}.

In this work, we propose a novel frame design for bistatic downlink ISAC where both communications and sensing aspects are simultaneously considered, which can be considered as a type of joint design. 
When the ISAC operates within a communication network, the typical solution is to perform sensing tasks using communication waveforms, such as \ac{OFDM}, \ac{OTFS}, and others.
In \cite{Sturm,Barneto}, the authors show the feasibility of performing radar processing using OFDM signals.
More recently, many works have considered OTFS due to its delay-Doppler (DD) domain processing \cite{Gaudio,Keskin,Gong}.
While radar processing using these waveforms works well in monostatic sensing, it is more challenging in the bistatic setting because it can be infeasible for the radar to know the transmitted data.

Given the practical difficulties in performing sensing using the data blocks, a typical approach to realize sensing using the communication infrastructure is to make use of the known signals within the frame to perform sensing.
In the context of the IEEE 802.11 standard, this can be done using the preamble composed of the short training field and the \ac{CEF}. 
In \cite{Nuria80211}, the authors show how to perform sensing for the IEEE 802.11ad frame using the preamble in a monostatic mmWave sensing. 
More recently, the authors in \cite{SundeepJump} have demonstrated a practical bistatic sensing setup exploiting the Golay sequences used for channel estimation in IEEE 802.11ay.
In the 5th generation (5G) of communications systems, a similar approach can be taken by using the positioning reference signal (PRS).
The PRSs are time-frequency resources within the resource block structure of 5G that enables parameter estimation. 
An approach of using PRS to perform sensing is shown in \cite{WeiPRS} as a pilot signal in ISAC.

When the pilot-based ISAC is designed specifically for sensing as in the PRS case, or when CEF is made more frequent in the IEEE 802.11 frame to improve sensing, the available data rate for communication is decreased.
Motivated by this problem, in this work we propose to include a deterministic signal in the guard interval instead of the \ac{CP}, such that the sensing signal does not compete with the data resources.
This approach is known in the literature to aid channel estimation \cite{Deneire,Coon,Huemer,ShahabTWC}, and is termed as a \ac{UW} scheme.
In \cite{Deneire}, the authors introduce the idea of replacing the CP with the UW to improve the synchronization for single-carrier (SC) systems.
In \cite{Coon}, the authors show how to track the time-varying channel using the UWs for SC, while the work \cite{Huemer} focuses on phase-tracking.
\rev{In \cite{ShahabTWC}, we have shown that the UW-based frame design achieves higher spectral efficiency than the CP-based systems for \ac{MIMO} \ac{GFDM} and OFDM.
Moreover, when the CP is removed or replaced by the UW, the CP-restoration technique \cite{BomfinTWC,BomfinCPfree} can be utilized to reconstruct the CP at the receiver.
That is, although removing or replacing the CP breaks the circular convolution structure of the channel, the frequency-domain equalization remains very similar to the CP transmission.}

In this paper, we combine the DD processing techniques for parameter estimation with the UW-based frame design and propose a new frame for ISAC, \revv{which is partially presented in \cite{Bomfin_Globecom}}.
The UW-based frame is advantageous for two reasons, namely, it does not require the radar receiver to know the communication data, and it does not incur data rate loss for communication.
Specifically, we design two UW signals based on the well-known \ac{ZC} sequence.
The first scheme has the same size as the CP and is termed UW1.
To enable a relatively simple \ac{FFT}-based DD estimation, we use the CP-restoration technique \cite{BomfinCPfree}, which leads to a loss in \ac{SNR}.
To avoid the SNR loss, a second scheme consisting of two ZC sequences with half the size of the CP is considered, which is termed UW2. The drawback of this second frame is that the estimation range is reduced.
However, we show that there can be cases where this reduction is not prohibitive, e.g., in the indoor environment.
\rev{Another advantage of these sequences is to enhance LoS synchronization similar to the approach of \cite{SundeepJump}, where the slow time-varying \ac{TO} and \ac{CFO} between the transmitter and radar nodes can be compensated.
There are several methods for TO and CFO cancellation available in the literature as shown by \cite{AndrewBistatic}. 
In this work, we address the TO and CFO compensation by an offset cancellation method in the delay-Doppler domain, since the TO and CFO are common to the LoS and target paths.}
Both UW1 and UW2 frames are compared to the OFDM and pilot-symbol (PS) benchmarks.
Here we note that OFDM has been chosen because it performs better than OTFS due to lower waterfall \ac{SNR}, as shown in \cite{Gaudio}.
For simplicity, we consider a PS approach that replaces a whole OFDM symbol as a pilot.
Additionally, we highlight that we consider a model with limited bandwidth based on the \ac{RRC} filter as \cite{Nuria80211}, which is typically not considered in other works in the literature.
In terms of theoretical results, we provide the \ac{CRB} that is valid for all frames, namely, UW1, UW2, OFDM, and PS.
\rev{The FFT-based multi-target estimator with LoS interference removal is derived and evaluated.}
Another useful theoretical result is the derivation of the outlier probability \ac{UB} in regards to the integer DD grid.
Differently from \cite{Gaudio}, our analysis considers the case where the true DD falls outside the grid search, yielding a novel expression. 
Lastly, we provide a comparative range analysis of the investigated frames.
The results reveal that both UW frames are competitive against the PS approach.
In particular, the UW1 frame is more suitable in outdoor deployment where the unambiguous range is broader.
In indoor deployments where the unambiguous range is not critical, the UW2 is more suitable, since it achieves the CRB.

In summary, the contributions are as follows
\begin{itemize}
\item Proposal of two UW-based frames for ISAC that do not require the knowledge of the data by the radar unit, and do not incur data rate reduction for communication. 
\item Derivation of the band-limited model with RRC filtering and its corresponding CRB.
\item \rev{Derivation of fine grid FFT-based multi-target DD estimator with LoS interference removal with TO and CFO cancellation.}
\item Derivation of the outlier probability UB for the integer grid FFT-based estimator. Two approximations for the UB are presented due to the numerical intractability of the exact solution.
\item Range comparison between the UW frames with the benchmarking frames, OFDM, and PS. The results show competitive performance in favor of UW.
\end{itemize}

 The remainder of the paper is organized as follows.
 In Section~\ref{sec:system_model}, the general system model and frames schemes are presented.
 In Section~\ref{sec:discrete_time_RX_signal}, the discrete-time channel model is shown with a suitable approximation that enables the FFT-based processing.
 In Section~\ref{sec:delay_doppler_estimation}, the CRB and DD estimators are derived.
 In Section~\ref{sec:outlier_probability}, the outlier probability expression and its approximations are presented.
 In Section~\ref{sec:frame_design_overview}, a frame design comparison is provided with all frames.
 The numerical results are shown in Section~\ref{sec:numerical_results}.
 Finally, Section~\ref{sec:conclusion} concludes the paper.

{{\it Notation}: Vectors and matrices are denoted by lowercase and uppercase bold symbols as $\mathbf{x}$ and $\mathbf{X}$, respectively, with indexing $\mathbf{x}[n]$ and $\mathbf{X}[n,m]$ returning the $n$th element of the vector, and the element in the $n$th row and $m$th column of the matrix.
The modulo operation is denoted as $(a)_N$. 
The transpose, conjugated transpose and conjugate are denoted by $(\cdot)^{\rm T}$, $(\cdot)^{\rm H}$ and $(\cdot)^\dagger$, respectively.
$\mathbf{A}\odot \mathbf{B}$ and $\mathbf{A}\oslash \mathbf{B}$ are the element-wise product and division between two equal-sized matrices.
The operations $(\mathbf{a} * \mathbf{b})[n]$ and $(\mathbf{a} \circledast \mathbf{b})[n]$ return the $n$th element of a linear and circular convolution between the vectors $\mathbf{a}$ and $\mathbf{b}$, which should be equal-sized in the latter case.
The normalized Fourier transform matrix of size $N$ is $\mathbf{F}_N$ where $\mathbf{F}_N\mathbf{F}_N^{\rm H} = \mathbf{I}_N$.
The expectation and variance operator are $\mathbb{E}(\cdot)$ and $\mathbb{V}(\cdot)$, respectively.
The real part of a complex number $a$ is $\mathcal{R}(a)$.
\rev{For a real number $a$, ${\rm round}(a) \in \mathbb{Z}$ returns the nearest integer to $a$.
A real random variable $x$ uniformly distributed between $[a,b]$ is denoted as $x\sim \mathcal{U}(a,b)$.}}

\section{System Model}\label{sec:system_model}
\begin{figure*}[t!]
	\centering
	\input{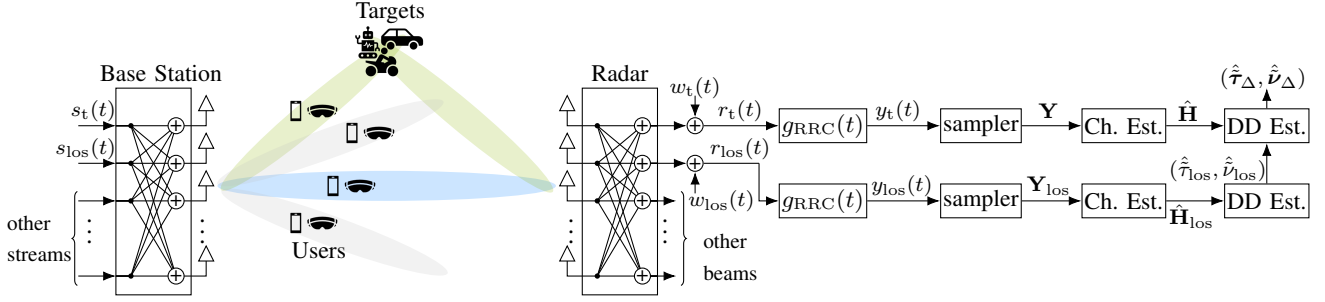}
	\vspace{-0.0cm}
	\caption{\rev{Bistatic radar system model using communications signals. The received signals $y_{\rm t}$ and $y_{\rm los}(t)$ are described in Section \ref{sec:system_model}. The sampled signals $\mathbf{Y}$ and $\mathbf{Y}_{\rm los}$ are given in Section \ref{sec:discrete_time_RX_signal}. The channel estimates $\hat{\mathbf{H}}$ and $\hat{\mathbf{H}}_{\rm los}$, and DD estimators are described in Section \ref{sec:delay_doppler_estimation}.}}
	\label{fig:system_model_high_level}
	\vspace{-0.2cm}
\end{figure*}

In this work, we consider a bistatic radar system operating within a wireless network that serves communication users in the downlink.
It is assumed that there is a line-of-sight (LoS) \rev{beam} between the base station (BS) and radar, such that the target's parameter estimation is performed in relation to the LoS component. 
An illustration of the system is shown in Fig.~\ref{fig:system_model_high_level}.
In the following, four different signaling schemes are considered.
The benchmarking schemes are OFDM and PS, while the proposed frames are UW1 and UW2.
%
\subsection{Transmit Signals with Different Frames}\label{subsec:tx_signals_frames}
In this work, we focus on the delay-Doppler processing, hence the spatial parameters such as angle-of-arrival and angle-of-departure are abstracted in the model for simplicity\footnote{\revv{In general the BS does not know the targets' positions. Thus, the beam which coincidentally hits a set of targets is processed by the radar, as illustrated by Fig.~\ref{fig:system_model_high_level}.}}.
The transmit signal \rev{of an arbitrary spatial stream} is composed of the concatenation of $M$ sub-blocks of $K+N_{\rm cp}$ samples each as 
\begin{equation}
	{\mathbf{x} = \left[ \tilde{\mathbf{x}}_{0}^{\rm T} \,\,\, \tilde{\mathbf{x}}_{1}^{\rm T} \,\,\,\cdots \,\,\, \tilde{\mathbf{x}}_{M-1}^{\rm T} \right]^{\rm T}},
\end{equation}
where ${\tilde{\mathbf{x}}_{m}} \in \mathbb{C}^{K+N_{\rm cp}}$ is the $m$th sub-block and ${\mathbf{x}} \in \mathbb{C}^{N_{\rm x}}$ is the concatenation of all sub-blocks and $N_{\rm x} = M(K+N_{\rm cp})$. 
The quantities $K$ and $N_{\rm cp}$ are the \ac{FFT} and CP sizes, respectively.
Once the discrete-time signal is generated, the continuous-time signal is generated by filtering using the RRC filter $g_{\rm RRC}(t)$ as
\begin{equation}
	{s}(t) = \sum_{n=0}^{N_{\rm x}-1} {\mathbf{x}}[n]g_{\rm RRC}(t-nT_{\rm s}),
\end{equation}
where $T_{\rm s}$ is the sampling interval, which is related to the bandwidth as $B=1/T_{\rm s}$.
\rev{In this work, we consider two spatial streams, namely, one associated with the LoS whose signals are termed $s_{\rm los}(t)$ and $\mathbf{x}_{\rm los}$, and ones associated with the targets are termed $s_{\rm t}(t)$ and $\mathbf{x}_{\rm t}$. 
We highlight that this notation has been chosen to derive the model with respect to the radar receiver and in practice $s_{\rm los}(t)$ and $s_{\rm t}(t)$ are seen as two arbitrary spatial streams from the base station viewpoint.}

\begin{figure}[t!]
	\centering
\begin{tikzpicture}[
	scale=0.9,
	>=stealth, 
	node distance=1.5cm, 
	block/.style={draw, rectangle, minimum height=0.6cm, minimum width=1cm}	
	]

	\node[above] at (-1.15,-0.25) {\small OFDM};
	\draw[-,dashed,gray] (-1,-0.5)--(7,-0.5) {};	

	\node[above] at (-1.35,-1.5) {\small PS};
	\draw[-,dashed,gray] (-1,-1.8)--(7,-1.8) {};	
	
	\node[above] at (-1.2,-2.7) {\small UW1};
	\draw[-,dashed,gray] (-1,-3.1)--(7,-3.1) {};	
	
	\node[above] at (-1.2,-3.8) {\small UW2};
	
	
	\node [block] (A) at (2,0) {\small $\mathbf{F}^{\rm H}_K$};
	\node [block, right of=A, node distance=1.6cm] (B) at (2,0) {CP};
	\node [block, right of=B, node distance=1.8cm] (C) at (3.7,0) {\small $g_{\rm RRC}(t)$};
	
	\draw[->,line width=0.8pt] (1,0) -- (A.west) node[pos=0.5, above] {\small $\mathbf{d}_m$};
	
	\draw[->,line width=0.8pt] (A) -- (B) node[midway, above] {\small $\mathbf{x}_m$};
	\draw[->,line width=0.8pt] (B) -- (C) node[midway, above] {\small $\tilde{\mathbf{x}}_m$};
	
	\draw[->] (C.east) -- ++(0.7,0) node[midway, above] {\small $s(t)$};	
	
	\newcommand{\deltay}{-1.25cm}
	\node [block] (Z) at (0.5,\deltay)  {\small Select};
	\node [block] (A) at (2,\deltay) {\small $\mathbf{F}^{\rm H}_K$};
	\node [block, right of=A, node distance=1.6cm] (B) at (2,\deltay) {\small CP};
	\node [block, right of=B, node distance=1.8cm] (C) at (3.7,\deltay) {\small $g_{\rm RRC}(t)$};
	
	\begin{scope}[transform canvas={yshift=0.2cm}]
	\draw[->,line width=0.8pt] (-0.5,\deltay) -- (Z.west) node[pos=0.5, above] {\small $\mathbf{d}_m$};
	\end{scope}
	
	\begin{scope}[transform canvas={yshift=-0.2cm}]
		\draw[->,line width=0.8pt] (-0.5,\deltay) -- (Z.west) node[pos=0.5, above,yshift=-0.05cm] {\small $\mathbf{p}_m$};
	\end{scope}
	
	\draw[->,line width=0.8pt] (Z) -- (A) node[midway, above] {};
	\draw[->,line width=0.8pt] (A) -- (B) node[midway, above] {\small $\mathbf{x}_m$};
	\draw[->,line width=0.8pt] (B) -- (C) node[midway, above] {\small $\tilde{\mathbf{x}}_m$};
	
	\draw[->] (C.east) -- ++(0.7,0) node[midway, above] {$s(t)$};

	\newcommand{\deltayuw}{-2.5cm}
	\node [block] (A) at (2,\deltayuw) {\small $\mathbf{F}^{\rm H}_K$};
	\node [block, right of=A, node distance=1.6cm] (B) at (2,\deltayuw) {UW1};
	\node [block, right of=B, node distance=1.8cm] (C) at (3.7,\deltayuw) {\small $g_{\rm RRC}(t)$};
	
	\draw[->,line width=0.8pt] (1,\deltayuw) -- (A.west) node[pos=0.5, above] {\small $\mathbf{d}_m$};
	
	\draw[->,line width=0.8pt] (A) -- (B) node[midway, above] {\small $\mathbf{x}_m$};
	\draw[->,line width=0.8pt] (B) -- (C) node[midway, above] {\small $\tilde{\mathbf{x}}_m$};
	
	\draw[->] (C.east) -- ++(0.7,0) node[midway, above] {$s(t)$};
	
	\newcommand{\deltayuww}{-3.75cm}
	\node [block] (A) at (2,\deltayuww) {\small $\mathbf{F}^{\rm H}_K$};
	\node [block, right of=A, node distance=1.6cm] (B) at (2,\deltayuww) {\small UW2};
	\node [block, right of=B, node distance=1.8cm] (C) at (3.7,\deltayuww) {\small $g_{\rm RRC}(t)$};
	
	\draw[->,line width=0.8pt] (1,\deltayuww) -- (A.west) node[pos=0.5, above] {\small$\mathbf{d}_m$};
	
	\draw[->,line width=0.8pt] (A) -- (B) node[midway, above] {\small$\mathbf{x}_m$};
	\draw[->,line width=0.8pt] (B) -- (C) node[midway, above] {\small$\tilde{\mathbf{x}}_m$};
	
	\draw[->] (C.east) -- ++(0.7,0) node[midway, above] {\small$s(t)$};

\end{tikzpicture}
	\caption{Block diagram of frame schemes \rev{of an arbitrary spatial stream}.}
	\vspace{-0.5cm}
	\label{fig:frames}
\end{figure}
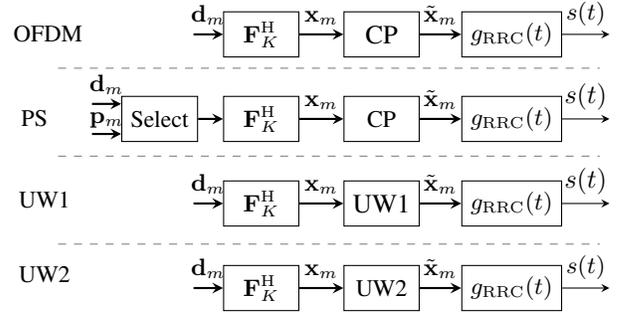

In the following, we show how $\tilde{\mathbf{x}}_m$ is generated by each transmission scheme considered in this manuscript.
A diagram is depicted in Fig.~\ref{fig:frames}.

\subsubsection{OFDM} 
the $m$th OFDM signal is 
\begin{equation}\label{eq:x_m_OFDM}
\mathbf{x}_m = \mathbf{F}_K^{\rm H}\mathbf{d}_m,
\end{equation}
where $\mathbf{d}_m$ is a \ac{QAM} symbol vector of size $K$ with $\mathbb{E}(\mathbf{d}_m \mathbf{d}_m^{\rm H}) = \mathbf{I}_K$ for all $m$.
The CP is added according to
\begin{equation}\label{eq:x_m_cp}
\tilde{\mathbf{x}}_m[n] = \mathbf{x}_m[(n-N_{\rm cp})_K],
\end{equation}
for $n = 0, 1, \cdots, K+N_{\rm cp}-1$ such that $\tilde{\mathbf{x}}_m \in \mathbb{C}^{K+N_{\rm cp}}$ has the first $N_{\rm cp}$ samples composed by the CP.

\subsubsection{PS}
In this case, the sub-blocks with indexes belonging to $\mathcal{M}_{\rm p} = \left\{0, M/M_{\rm p}, 2M/M_{\rm p}, \cdots, (M_{\rm p}-1)M/M_{\rm p}  \right\}$ with $M_{\rm p} = |\mathcal{M}_{\rm p}|$ elements are the PS $\mathbf{p}_m \in \{-1,1\}^{K}$, which is essentially a pilot signal with the same size as the data symbol.
The $m$th sub-block is given by
\begin{equation}
	\mathbf{x}_m =  \left\{\begin{matrix}
		\mathbf{F}_K^{\rm H}\mathbf{p}_m, & m\in \mathcal{M}_{\rm p}	
		\\ \mathbf{F}_K^{\rm H}\mathbf{d}_m	, & m\in \mathcal{M}_{\rm d}
	\end{matrix}\right. ,
\end{equation}
where the data set $\mathcal{M}_{\rm d}$ is disjoint to $\mathcal{M}_{\rm p}$ such that $\mathcal{M}_{\rm p} \cup \mathcal{M}_{\rm d} = \left\{0,1,\cdots, M-1 \right\}$.
The CP is added similarly to \eqref{eq:x_m_cp}.

\subsubsection{UW1}
The UW1 frame has the OFDM waveform in all sub-blocks according to \eqref{eq:x_m_OFDM}, but instead of CP, the deterministic signal $\mathbf{x}_{\rm uw} \in \mathbb{C}^{N_{\rm cp}}$ is appended. 
The UW1 is by the $N_{\rm cp}$ size ZC sequence as
\begin{equation}\label{eq:x_m_UW}
\mathbf{x}_{\rm uw1}[n] = \exp(-j\pi n^2/N_{\rm cp}),
\end{equation}
for $0\leq n < N_{\rm cp}$.
Then the $m$th sub-block $\tilde{\mathbf{x}}_m \in \mathbb{C}^{K+N_{\rm cp}}$ with UW1 transmission is given by
\begin{equation}
	\tilde{\mathbf{x}}_m = \left[ \mathbf{x}_{\rm uw1}^{\rm T} \,\,\, {\mathbf{x}}_m^{\rm T}\right]^{\rm T}.
\end{equation}
\subsubsection{UW2}
The UW2 transmission has a similar concept to the UW1.
The difference is that the UW2 signal $\mathbf{x}_{\rm uw2} \in \mathbb{C}^{N_{\rm cp}/2}$ is half the size of the ZC signal \eqref{eq:x_m_UW} as
\begin{equation}\label{eq:x_uw2}
	\mathbf{x}_{\rm uw2}[n] = \exp(-j\pi n^2/(N_{\rm cp}/2)),
\end{equation}
for $0\leq n < N_{\rm cp}/2$, so that it needs to be appended twice before the OFDM symbol.
Then the $m$th sub-block $\tilde{\mathbf{x}}_m \in \mathbb{C}^{K+N_{\rm cp}}$ with UW2 transmission is given by
\begin{equation}
	\tilde{\mathbf{x}}_m = \left[ \mathbf{x}_{\rm uw2}^{\rm T} \,\,\, \mathbf{x}_{\rm uw2}^{\rm T} \,\,\, {\mathbf{x}}_m^{\rm T}\right]^{\rm T}.
\end{equation}
The idea of using a ZC sequence with half the size in \eqref{eq:x_uw2} is to allow a simpler radar receiver and to not incur interference as the one in \eqref{eq:x_m_UW}.
As it will be shown in Section \ref{sec:frame_design_overview}, the UW2 signaling has half of the maximum unambiguous delay when the FFT-based radar receiver is used.
Nevertheless, we note that it is possible to design a more sophisticated radar receiver that expands the maximum delay, which is not explored in this work.

\begin{figure}[t!]
	\centering
\begin{tikzpicture}[scale=0.7]
	
  \begin{axis}[
	axis lines=none,  
	ticks=none,       
	xmin=-5, xmax=5,
	ymin=-5, ymax=5,
	scale only axis,
	width=\textwidth,  
	height=\textwidth, 
	clip=false         
	]
	\addplot[color=azure(colorwheel),   line width=2pt,    smooth,mark=Mercedes star,only marks,mark size=7pt, mark options = {fill=azure(colorwheel)}] coordinates {(0,0)};
	
	\addplot[color=azure(colorwheel),   line width=2pt,    smooth,mark=Mercedes star flipped,only marks,mark size=7pt, mark options = {fill=azure(colorwheel)}] coordinates {(3,0)};
	
	\addplot [color=black, solid,   mark = x,   only marks,mark size = 5.8pt,line width = 1.75] coordinates {(3/2,1)};
	
	\addplot [black, dashed,  line width=0.5pt] 
	table[row sep=crcr]{
		0   0\\ 1.5 1\\ 3 0\\
	};
		\addplot [black, dashed,  line width=0.5pt,] 
	table[row sep=crcr]{
		0   0\\ 3 0\\
	};
	
\end{axis}
	
	\node [] at (9.075,8.6){BS};
	\node [] at (9.1+5.40,8.6){R};
	\node [] at (9.1+5.40/2,11.3){T$_p$};
	
	\node [] at (9.1+5.35/2,8.75){$d_{{\rm BR}_p}$};
	\node [] at (9.1+4.1/4-0.2	,10.25){$d_{{\rm BT}_p}$};
	\node [] at (9.1+5.7/4*3+0.1,10.25){$d_{{\rm TR}_p}$};
	
	

\end{tikzpicture}
	\vspace{-0.2cm}
	\caption{\rev{Bistatic radar geometry.}}
	\label{fig:radar_geometry}
\end{figure}
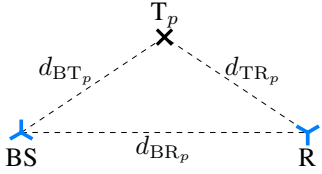
\begin{figure}[t!]
	\centering
	\input{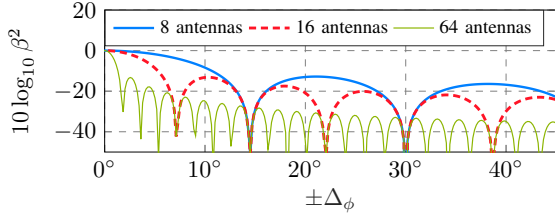}
	\vspace{-0.2cm}
	\caption{\rev{Example of normalized \ac{ULA} beam pattern that determines the LoS rejection for a given angle difference between the LoS and target beam.}}
	\label{fig:beam_pattern}
	\vspace{-0.3cm}
\end{figure}

\setcounter{mytempeqncnt}{\value{equation}}
\setcounter{equation}{12}
\begin{figure*}[b!]
	\rev{\hrulefill}
	\rev{\begin{equation}\label{eq:y_approx}
			y_{\rm t}(t)  \approx \sum_{p=0}^{P-1} h_p e^{-j 2 \pi \nu_p}	\sum_{n=0}^{N_{\rm x}-1} \mathbf{x}_{\rm t}[n]g_{\rm RC}(t-nT_{\rm s}-\tau_p) 
			+	\beta \left(h_{\rm los} e^{-j 2 \pi \nu_{\rm los}}\sum_{n=0}^{N_{\rm x}-1} \! \mathbf{x}_{\rm los}[n]g_{\rm RC}(t\!-\! nT_{\rm s}\!-\!\tau_{\rm los})\right)	 + w_{\rm t}'(t).
	\end{equation}}
	\rev{\begin{equation}\label{eq:y_los_approx}
			y_{\rm los}(t) \approx h_{\rm los} e^{-j 2 \pi \nu_{\rm los}}\sum_{n=0}^{N_{\rm x}-1} \! \mathbf{x}_{\rm los}[n]g_{\rm RC}(t\!-\! nT_{\rm s}\!-\!\tau_{\rm los})
			+\beta\left(\sum_{p=0}^{P-1} h_p e^{-j 2 \pi \nu_p}	\sum_{n=0}^{N_{\rm x}-1} \mathbf{x}_{\rm t}[n]g_{\rm RC}(t-nT_{\rm s}-\tau_p)\right) + w_{\rm los}'(t) 
	\end{equation}}
\end{figure*}
\setcounter{equation}{\value{mytempeqncnt}}

\subsection{Bistatic Radar Channel}
%
\rev{In this work, we consider a scenario with $P$ targets.
	An exemplary radar geometry is given in Fig.~\ref{fig:radar_geometry} for an arbitrary target with index $p \in \left\{0,1,\cdots, P-1\right\}$, where the distances in meters between the BS and target, target and radar, and BS and radar are denoted as $d_{{\rm BT}_p}$, $d_{{\rm TR}_p}$ and $d_{{\rm BR}_p}$, respectively.
The time delay associated with each path is given by $\tau_{{\rm BT}_p}=d_{{\rm BT}_p}/c$, $\tau_{{\rm TR}_p}=d_{{\rm TR}_p}/c$ and $\tau_{{\rm BR}_p}=d_{{\rm BR}_p}/c$, where $c$ is the speed of light.
We consider the \ac{TO} and \ac{CFO} due to the asynchronicity between the BS and the radar to be unchanged during the \ac{CPI}, which are denoted by $\tau_{\rm to}$ and $\nu_{\rm cfo}$, respectively. 
\\
Based on the considerations above, the delay and frequency shift associated to the LoS are given by $\tau_{\rm los} = \tau_{\rm BT} + \tau_{\rm to}$ and $\nu_{\rm los}  = \nu_{\rm cfo}$, respectively.
The delay and frequency shift for the path associated to the $p$th target are $\tau_p = \tau_{{\rm BT}_p} + \tau_{{\rm TR}_p} + \tau_{\rm to}$ and $\nu_p = f_p + \nu_{\rm cfo}$, where $f_p$ is the frequency-shift due to Doppler.
Thus, the channel impulse response associated to the LoS path is $h_{\rm los}(t,\tau) =  h_{\rm los}\delta(\tau-\tau_{\rm los}) e^{-j 2 \pi \nu_{\rm los} t}$ with the channel gain $h_{\rm los}$.
And the radar channel impulse response for the $p$th target is $h_p(t,\tau) =  h_p\delta(\tau-\tau_p) e^{-j 2 \pi \nu_p t}$ with the channel gain $h_p$.
In practice, the channel gains $|h_p|^2$ and $|h_{\rm los}|^2$ are determined by the well-known radar double-path and path-loss formulas, as considered in the high-level analysis of Subsection \ref{subsec:high_level_analysis}.}
\rev{The phase $\angle h_p$ and $\angle h_{\rm los}$ are assumed to be unchanged during the \ac{CPI} but random with a uniform distribution between 0 and $2\pi$ for each new frame and independent from one another.}

\rev{We note that ultimately we are interested in estimating the difference $\tau_p - \tau_{\rm los}$ and $\nu_p - \nu_{\rm los}$ to infer the position of the target, where the TO and CFO effects are compensated since they are common for the LoS and target paths. This method can be categorized as one type of offset cancellation method in the delay-Doppler domain \cite{AndrewBistatic}.}


\subsection{Received Signals}
\subsubsection{Paths Associated to Targets}
Assuming that the radar channel stays constant during the transmission of $s_{\rm t}(t)$, the received signal associated with the radar path after beamforming is given by
\begin{equation}\label{eq:r}
\rev{\begin{split}
	r_{\rm t}(t)  = & \sum_{p=0}^{P-1} h_p s_{\rm t}(t-\tau_p) e^{-j 2 \pi \nu_p t} \\ & + \beta h_{\rm los} s_{\rm los}(t-\tau_{\rm los}) e^{-j 2 \pi \nu_{\rm los} t} + w_{\rm t}(t),
	\end{split}}
\end{equation}
\rev{where $\beta \in [0,1]$ represents the LoS rejection due to beamfoming. For example, if the radar employs an \ac{ULA} with 64 antennas, $\beta^2 = 1/100$ corresponding to a $20\,{\rm dB}$ of rejection is obtained if the LoS and the target are more than 5 degrees apart, i.e., $|\Delta_\phi| > 5^\circ$ in Fig.~\ref{fig:beam_pattern}.} Lastly, $w(t)$ is the \ac{AWGN}.
\revv{The model of \eqref{eq:r} assumes that each target has one significant reflection to the radar.} 

The received signal is filtered using the same root-raised cosine filter as the transmitter
\begin{equation}\label{eq:y}
	\rev{y_{\rm t}(t) = \int r_{\rm t}(t') g_{\rm RRC}(t-t')dt'}.
\end{equation}
\rev{An approximation is given in equation \eqref{eq:y_approx} at the bottom of this page which is used in the simulation to simplify the model.}
\rev{This approximation is typically considered in the literature and assumes that the bandwidth of $g_{\rm RRC}(t)$ is much larger than the maximum frequency shift.}
Intuitively, this means that the linear phase shift imposed by the channel is negligible within the time duration of the filter $g_{\rm RRC}(t)$. 
In \eqref{eq:y}, $g_{\rm RC}(t)$ is the raised cosine filter	
\begin{align}
g_{\rm RC}(t) = \left\{\begin{matrix}
	\frac{\pi}{4T_{\rm s}} {\rm sinc}\left(\frac{1}{2\alpha}\right), & t = \pm T_{\rm s}/(2\alpha)
	\\ 	\frac{1}{T_{\rm s}} {\rm sinc} \left(\frac{1}{T_{\rm s}} \right)	\frac{\cos(\pi \alpha t/T_{\rm s})}{1-(2\alpha t/T_{\rm s})^2}, & {\rm otherwise}
\end{matrix}\right.
\end{align}
with roll-off factor $\alpha$.
\subsubsection{Path Associated to the LoS}
\rev{Analogously to \eqref{eq:y_approx}, assuming that the radar channel stays constant during the transmission of $s_{\rm los}(t)$, the approximate received signal associated with the LoS path is given in equation \eqref{eq:y_los_approx} at the bottom of the previous page.}
\section{Discrete-Time Received Signal}\label{sec:discrete_time_RX_signal}
In the following, we describe how the passive radar samples $y_{\rm t}(t)$ in \eqref{eq:y_approx} for each transmit signal of Subsection \ref{subsec:tx_signals_frames}.
Additionally, a simplified model is derived, which is then used to design the DD estimator in Section \ref{sec:delay_doppler_estimation}.
\rev{Although this section focuses on the targets branch of Fig.~\ref{fig:system_model_high_level}, the sampled signal associated with the LoS branch can be obtained by sampling $y_{\rm los}(t)$ in \eqref{eq:y_los_approx} and following an analogous procedure.}
\subsection{\rev{OFDM Frame}} 
\rev{In this case, the passive radar uses all OFDM sub-blocks to perform parameter estimation.
As shown in the Appendix \ref{ap_subsec:discrete_time_signal}, the received signal sampled at the sampling interval $T_{\rm s}$ is approximated as}
\setcounter{equation}{14}
\begin{align}\label{eq:Y_ofdm}
	& \rev{\mathbf{Y}_{\rm ofdm}[k,m] = y_{\rm t}(I_{\rm ofdm}(k,m)T_{\rm s})} \nonumber
	 \\ & \rev{\approx \sum_{p=0}^{P-1}h_p' \mathbf{u}_{M,\tilde{\nu}_p}[m] (\mathbf{x}_{{\rm t}_m} \circledast \mathbf{g}_{K,\tilde{\tau}_p})[k]}
\nonumber\\ &  \rev{+\beta h_{\rm los}' \mathbf{u}_{M,\tilde{\nu}_{\rm los}}[m] (\mathbf{x}_{{\rm los}_m} \circledast \mathbf{g}_{K,\tilde{\tau}_{\rm los}})[k]+ \mathbf{W}_{\rm ofdm}[k,m]},
\end{align}
\rev{where $I_{\rm ofdm}(k,m)$ is the index mapper, $\mathbf{g}_{K,\tilde{\tau}_p}[k] = \tilde{g}_{\rm RC}(k-\tilde{\tau}_p)$ is the discrete-time raised cosine filter for $k=0,1,\cdots,K-1$ and $\tilde{g}_{\rm RC}(t) = {g}_{\rm RC}(t_p/T_{\rm s})$, $\tilde{\tau}_p = \tau_p/T_{\rm s}$ is the delay normalized to the sampling interval, which is in the interval $0\leq \tilde{\tau}_p < N_{\rm cp}$.
We consider that the normalized LoS delay $\tilde{\tau}_{\rm los} = \tau_{\rm los}/T_{\rm s}$ is in the interval $0\leq \tilde{\tau}_{\rm los} < 1$, i.e., the integer delay of the LoS path is in the first sample of the channel impulse response, which assumes that the system is synchronized with respect to the LoS.
The frequency shift related component is $\mathbf{u}_{M,\tilde{\nu}_p}[m] = \exp(-j 2 \pi \tilde{\nu}_pm/M)$, where $\tilde{\nu}_p = \nu_p T_{\rm s} N_{\rm x}$ is the frequency shift normalized to subcarrier spacing of the whole signal if the bandwidth is divided by $N_{\rm x}$.
The noise term $\mathbf{W}_{\rm ofdm}[k,m]$ has power $\sigma_{\rm w}^2$ and is uncorrelated for different $k$ and/or $m$.}

\rev{Lastly, we note that for the OFDM frame, the spatial streams associated with the target and LoS have different OFDM symbols, which are distinguished in \eqref{eq:Y_ofdm} by $\mathbf{x}_{{\rm t}_m}$ and $\mathbf{x}_{{\rm los}_m}$, respectively.}
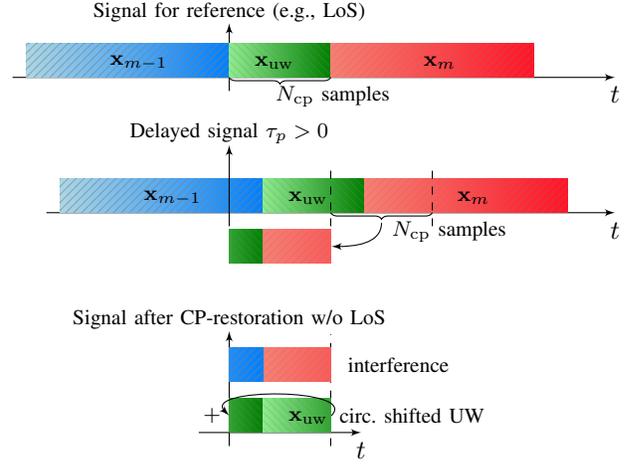
\begin{figure}[t!]
	\centering
	\tikzstyle{block} = [draw, fill=white, rectangle, 
minimum height=3em, minimum width=3em]
\tikzstyle{block2} = [draw, fill=white, rectangle, 
minimum height=1em, minimum width=2.4em]

\tikzstyle{multiplier} = [draw,circle,fill=blue!20,add={}{}{}{}] {} 
\tikzstyle{sum} = [draw,circle,scale=0.7,add2={}{}{}{}] {} 
\tikzstyle{input} = [coordinate]
\tikzstyle{output} = [coordinate]
\tikzstyle{pinstyle} = [pin edge={to-,thin,black}]
\newcommand\z{3}

\def\windup{
	\tikz[remember picture,overlay]{
		\draw (-0.8,0) -- (0.8,0);
		\draw (-0.8,-0.4)--(-0.4,-0.4) -- (0.4,0.4) --(0.8,0.4);
}}

\usetikzlibrary{matrix,decorations.pathreplacing,calc,positioning,calligraphy}

\centering
\begin{tikzpicture}[auto, node distance=0cm,>=latex',scale=0.9]
	
	\def\delta_y{1/2};
	\def\x_ref{3};
	\def\uw_size{1.5};
	\def\uwdelay{0.5};
	
	\draw[->] (\x_ref, -1/5) -- (\x_ref, 0.8) node[above,yshift=-0.1cm] {\footnotesize Signal for reference (e.g., LoS)};	
	
	\draw[->] (-0.2, 0) -- (8.7, 0) node[below] {$t$};

	\shade[left color=lightblue,right color=azure(colorwheel)] (0,0) -- (\x_ref,0) -- (\x_ref,\delta_y) -- (0,\delta_y); 
	\pattern[pattern=north east lines, pattern color=gray,opacity=0.75,dashed] (0,0) rectangle (\x_ref,\delta_y); 
	\node[xshift = 1.5cm, yshift = 0.20cm, opacity = 1]{\footnotesize $\mathbf{x}_{m-1}$};
	
	\shade[left color=lightgreen,right color=mediumgreen] (\x_ref,0) -- (\x_ref+\uw_size,0) -- (\x_ref+\uw_size,\delta_y) -- (\x_ref,\delta_y);
	\pattern[pattern=north west lines, pattern color=gray,opacity=0.5] (\x_ref,0) rectangle (\x_ref+\uw_size,\delta_y); 
	\node[xshift = 3.3cm, yshift = 0.20cm, opacity = 1]{\footnotesize $\mathbf{x}_{\rm uw}$};
	
	\shade[left color=salmon,right color=awesome] (\x_ref+\uw_size,0) -- (\x_ref+\uw_size+\x_ref,0) -- (\x_ref+\uw_size+\x_ref,\delta_y) -- (\x_ref+\uw_size,\delta_y);
	\pattern[pattern=north east lines, pattern color=gray,opacity=0.5,loosely dashed] (\x_ref+\uw_size,0) rectangle (\x_ref+\uw_size+\x_ref,\delta_y); 
	\node[xshift = 5.5cm, yshift = 0.20cm, opacity = 1]{\footnotesize $\mathbf{x}_{m}$};
	
			\draw[decorate, decoration={brace, amplitude=2.5pt}] (4.5,-0.0) -- (3,-0.0) 
	node[midway, below=4pt,align=left,xshift=0.75cm,yshift=0.15cm] {\footnotesize $N_{\rm cp}$ samples };
	
		\begin{scope}[shift={(\uwdelay,-2)}]
			\draw[->] (-0.2, 0) -- (8.2, 0) node[below] {$t$};
			\shade[left color=lightblue,right color=azure(colorwheel)] (0,0) -- (\x_ref,0) -- (\x_ref,\delta_y) -- (0,\delta_y); 
		\pattern[pattern=north east lines, pattern color=gray,opacity=0.75,dashed] (0,0) rectangle (\x_ref,\delta_y); 
		\node[xshift = 1.5cm, yshift = 0.20cm, opacity = 1]{\footnotesize $\mathbf{x}_{m-1}$};
		
		\shade[left color=lightgreen,right color=mediumgreen] (\x_ref,0) -- (\x_ref+\uw_size,0) -- (\x_ref+\uw_size,\delta_y) -- (\x_ref,\delta_y);
		\pattern[pattern=north west lines, pattern color=gray,opacity=0.5] (\x_ref,0) rectangle (\x_ref+\uw_size,\delta_y); 
		\node[xshift = 3.3cm, yshift = 0.20cm, opacity = 1]{\footnotesize $\mathbf{x}_{\rm uw}$};
		
		\shade[left color=salmon,right color=awesome] (\x_ref+\uw_size,0) -- (\x_ref+\uw_size+\x_ref,0) -- (\x_ref+\uw_size+\x_ref,\delta_y) -- (\x_ref+\uw_size,\delta_y);
		\pattern[pattern=north east lines, pattern color=gray,opacity=0.5,loosely dashed] (\x_ref+\uw_size,0) rectangle (\x_ref+\uw_size+\x_ref,\delta_y); 
		\node[xshift = 5.5cm, yshift = 0.20cm, opacity = 1]{\footnotesize $\mathbf{x}_{m}$};		
		\end{scope}
		
		\draw[->] (\x_ref, -2.2) -- (\x_ref, -1) node[above,align = left,yshift=-0.1cm] {\footnotesize Delayed signal $\tau_p>0$};

		\begin{scope}[shift={(-\uw_size+\uwdelay,-2.75)}]
			\clip (3*\uw_size-\uwdelay,0) rectangle (4*\uw_size-\uwdelay,\delta_y); 
			\draw[->] (-0.2, 0) -- (8.2, 0) node[below] {$t$};
			
		\shade[left color=lightgreen,right color=mediumgreen] (\x_ref,0) -- (\x_ref+\uw_size,0) -- (\x_ref+\uw_size,\delta_y) -- (\x_ref,\delta_y);
		\pattern[pattern=north west lines, pattern color=gray,opacity=0.5] (\x_ref,0) rectangle (\x_ref+\uw_size,\delta_y); 
		\node[xshift = 3.3cm, yshift = 0.20cm, opacity = 1]{\footnotesize $\mathbf{x}_{\rm uw}$};
		
		\shade[left color=salmon,right color=awesome] (\x_ref+\uw_size,0) -- (\x_ref+\uw_size+\x_ref,0) -- (\x_ref+\uw_size+\x_ref,\delta_y) -- (\x_ref+\uw_size,\delta_y);
		\pattern[pattern=north east lines, pattern color=gray,opacity=0.5,loosely dashed] (\x_ref+\uw_size,0) rectangle (\x_ref+\uw_size+\x_ref,\delta_y); 
		\node[xshift = 5.5cm, yshift = 0.20cm, opacity = 1]{\footnotesize $\mathbf{x}_{m}$};
		\end{scope}
		\draw[-,dashed] (\x_ref+\uw_size, -2.2) -- (\x_ref+\uw_size, -1.3);	
		\draw[-,dashed] (\x_ref+2*\uw_size, -2.2) -- (\x_ref+2*\uw_size, -1.3);	
		
		\draw[decorate, decoration={brace, amplitude=2.5pt}] (6,-2.0) -- (4.5,-2.0) 
		node[midway, below=4pt,align=left,xshift=0.95cm,yshift=0.15cm] {\footnotesize $N_{\rm cp}$ samples };
		
		\draw [->] (5.25,-2.15) to [out=-90,in=0] (4.51,-2.5);

		\draw[->] (\x_ref, -5.2-0.25) -- (\x_ref, -3.75) node[above,yshift=-0.10cm] {\footnotesize Signal after CP-restoration w/o LoS};	
		\draw[->] (-0.2+2.75, -5-0.25) -- (8.2-3.25, -5-0.25) node[below] {$t$};
		\draw[-,dashed] (\x_ref+\uw_size, -5.2-0.25) -- (\x_ref+\uw_size, -3.8);	

		\begin{scope}[shift={(-\uw_size+\uwdelay,-5.25)}]
			\clip (3*\uw_size-\uwdelay,0) rectangle (4*\uw_size-\uwdelay,\delta_y); 

			\shade[left color=lightgreen,right color=mediumgreen] (\x_ref,0) -- (\x_ref+\uw_size,0) -- (\x_ref+\uw_size,\delta_y) -- (\x_ref,\delta_y);
			\pattern[pattern=north west lines, pattern color=gray,opacity=0.5] (\x_ref,0) rectangle (\x_ref+\uw_size,\delta_y); 
			\node[xshift = 3.3cm, yshift = 0.20cm, opacity = 1]{\footnotesize $\mathbf{x}_{\rm uw}$};			
		\end{scope}
		
				\begin{scope}[shift={(\uwdelay,-5.25)}]
			\clip (\uwdelay+\uw_size,0) rectangle (3*\uw_size-\uwdelay,\delta_y); 
			
			\shade[left color=lightgreen,right color=mediumgreen] (\x_ref,0) -- (\x_ref+\uw_size,0) -- (\x_ref+\uw_size,\delta_y) -- (\x_ref,\delta_y);
			\pattern[pattern=north west lines, pattern color=gray,opacity=0.5] (\x_ref,0) rectangle (\x_ref+\uw_size,\delta_y); 
			\node[xshift = 3.3cm, yshift = 0.20cm, opacity = 1]{\footnotesize $\mathbf{x}_{\rm uw}$};			
		\end{scope}
		
				\draw [->] (4.51,-5.0) to [out=45,in=135] (3.0,-5.0);
		\node[align=left] at (5.7,-5) {\footnotesize circ. shifted UW};
		\node[align=left] at (2.75,-5) {\footnotesize $+$};
		
		\begin{scope}[shift={(-\uw_size+\uwdelay,-4.5)}]
			\clip (3*\uw_size-\uwdelay,0) rectangle (4*\uw_size-\uwdelay,\delta_y); 
			
			\shade[left color=salmon,right color=awesome] (\x_ref+\uw_size,0) -- (\x_ref+\uw_size+\x_ref,0) -- (\x_ref+\uw_size+\x_ref,\delta_y) -- (\x_ref+\uw_size,\delta_y);
		\pattern[pattern=north east lines, pattern color=gray,opacity=0.5,loosely dashed] (\x_ref+\uw_size,0) rectangle (\x_ref+\uw_size+\x_ref,\delta_y); 
		\end{scope}
		
				\begin{scope}[shift={(\uwdelay,-4.5)}]
			\clip (\x_ref-\uwdelay,0) rectangle (\x_ref,\delta_y);
			
			\shade[left color=lightblue,right color=azure(colorwheel)] (0,0) -- (\x_ref,0) -- (\x_ref,\delta_y) -- (0,\delta_y); 
			\pattern[pattern=north east lines, pattern color=gray,opacity=0.75,dashed] (0,0) rectangle (\x_ref,\delta_y); 
		\end{scope}
		
		\node[align=left] at (5.5,-4.25) {\footnotesize interference};

\end{tikzpicture}
\vspace{-0.3cm}
	\caption{\rev{Illustration of CP-restoration.}}
	\label{fig:uw_cp_restoration}
\end{figure}

\begin{figure*}[b!]
	\setcounter{mytempeqncnt}{\value{equation}}
	\setcounter{equation}{23}
	\rev{\hrulefill}
	\rev{\begin{equation}\label{eq:Y_uw}
			\begin{split}
				\mathbf{Y}_{\rm uw1}[k,m]  & =  y_{\rm t}(I_{\rm uw1}(k,m)T_{\rm s})  + \underbrace{y_{\rm t}((I_{\rm uw1}(k,m)+N_{\rm cp})T_{\rm s})}_{\rm CP-restoration}
				\\ & \approx \sum_{p=0}^{P-1}h_p' \mathbf{u}_{M,\tilde{\nu}_p}[m] (\mathbf{x}_{\rm uw1} \circledast \mathbf{g}_{N_{\rm cp},\tilde{\tau}_p})[k] + \beta h_{\rm los}' \mathbf{u}_{M,\tilde{\nu}_{\rm los}}[m] (\mathbf{x}_{\rm uw1} \circledast \mathbf{g}_{K,\tilde{\tau}_{\rm los}})[k] + \mathbf{W}_{\rm uw1}[k,m]
			\end{split} 
	\end{equation}	}
	\rev{\begin{equation}\label{eq:W_uw}
			\begin{split}
				\mathbf{W}_{\rm uw1}[k,m] & \approx  w_{\rm t}'(I_{\rm uw1}(k,m)T_{\rm s}) + \underbrace{w_{\rm t}'((I_{\rm uw1}(k,m)+N_{\rm cp})T_{\rm s})}_{\rm CP-restoration \, AWGN} + \underbrace{\beta h_{\rm los}' \mathbf{u}_{M,\tilde{\nu}_{\rm los}}[m](\mathbf{x}_m * \mathbf{g}_{K,\tilde{\tau}_{\rm los}})[k]}_{{\rm LoS \, interference \, from} \, m{\rm th \,  block}} 
				\\ & + \sum_{p=0}^{P-1}\underbrace{ h_p' \mathbf{u}_{M,\tilde{\nu}_p}[m-1] (\mathbf{x}_{m-1} * \mathbf{g}_{K,\tilde{\tau}_p})[k+K]}_{{\rm interference \, from} \, (m-1){\rm th \,  block}} 
				+ \underbrace{h_p' \mathbf{u}_{M,\tilde{\nu}_p}[m](\mathbf{x}_m * \mathbf{g}_{K,\tilde{\tau}_p})[k]}_{{\rm interference \, from} \, m{\rm th \,  block}}
			\end{split}
	\end{equation}}
	\setcounter{equation}{\value{mytempeqncnt}}
\end{figure*}

\subsection{PS Frame} 
When the passive radar performs the processing only based on the PS signals, the sampled signal is $\mathbf{Y}_{\rm ps} \in \mathbb{C}^{K\times M_{\rm p}}$ defined as
\setcounter{equation}{15}
\begin{align}\label{eq:Y_prs}
		&\rev{\mathbf{Y}_{\rm ps}[k,m] \approx \sum_{p=0}^{P-1}h_p' \mathbf{u}_{M_{\rm p},\tilde{\nu}_p}[m] (\mathbf{x}_m \circledast \mathbf{g}_{K,\tilde{\tau}_p})[k]} \nonumber
		\\ & \rev{+ \beta h_{\rm los}' \mathbf{u}_{M_{\rm p},\tilde{\nu}_{\rm los}}[m] (\mathbf{x}_m \circledast \mathbf{g}_{K,\tilde{\tau}_{\rm los}})[k] + \mathbf{W}_{\rm ps}[k,m M/M_{\rm p}],}
\end{align}	
for $0\leq k < K$ and $ 0\leq m < M_{\rm p}$, which has an analogous structure to $\mathbf{Y}_{\rm ofdm}$ in \eqref{eq:Y_ofdm} by following the same steps, but with columns that respect $m \in \mathcal{M}_{\rm p}$.
\rev{Also, the noise sample of $\mathbf{W}_{\rm ps}$ have the noise power of $\mathbf{W}_{\rm ofdm}$, $\sigma_{\rm w}^2$.}
It is important to notice that $\mathbf{u}_{M_{\rm p},\tilde{\nu}_l}$ has the same form as the one in \eqref{eq:Y_ofdm} but with $M_{\rm p}$ instead of $M$.
\rev{Another important distinction to \eqref{eq:Y_ofdm} is that the pilot symbols are assumed to be the same for all spatial streams, which leads to the same signal $\mathbf{x}_m$ for the target and LoS paths $\forall m \in \mathcal{M}_{\rm p}$.}
\subsection{UW1 Frame with CP-restoration} 
For the UW1 transmission scheme, a frequency domain processing can be accomplished by the radar after performing the CP-restoration process similar to \cite{BomfinTWC}. 
\rev{This process is illustrated in Fig.~\ref{fig:uw_cp_restoration}, which consists of adding the signals $N_{\rm cp}$ samples apart.
The received signal is described in equation \eqref{eq:Y_uw} at the bottom of the next page} for $0\leq k < N_{\rm cp}$ and $ 0\leq m < M$ such that $\mathbf{Y}_{\rm uw1} \in \mathbb{C}^{N_{\rm cp} \times M}$.
The index mapper function $I_{\rm uw1}(k,m) = (k+ m(N_{\rm cp} + K))$ ensures that the UW1 portion of the signal is sampled.
\rev{Neglecting the AWGN noise, Fig.~\ref{fig:uw_cp_restoration} demonstrates that the resulting signal is the cyclic convolved UW1 plus interference from the $(m-1)$th and $m$th OFDM signals.}
\rev{Naturally, extra noise samples are also copied, so the resulting noise plus interference term is given in \eqref{eq:W_uw} at the bottom of the next page and has power}
\begin{equation}\label{eq:noise_power_uw}
	\begin{split}
& \mathbb{E}(\mathbf{W}_{\rm uw1}[k,m]\mathbf{W}_{\rm uw1}[k,m]^{\rm H}) \approx \\ & \rev{2\sigma_{\rm w}^2 + \sum_{p=0}^{P-1}|h_p|^2 + \beta^2 |h_{\rm los}|^2= \sigma_{\rm w}^2 L_{\rm uw1}.}
\end{split}
\end{equation}
\rev{The AWGN power is counted twice due to the CP-restoration AWGN component in \eqref{eq:W_uw}.
The remaining terms are the interference from the $(m-1)$th and $m$th OFDM blocks.
It is important to notice that for a given time index $k$, either the $(m-1)$th or $m$th block has dominant power.
This fact is depicted in Fig.~\ref{fig:uw_cp_restoration} where the interference from the  $(m-1)$th and $m$th OFDM interference symbols are not superposed but occupy different time slots.
Thus, the power $|h_p|^2$ is added once and not twice.}
We can conveniently define
\begin{equation}\label{eq:L_loss_uw}
	\rev{L_{\rm uw1} = 2 + 1/\sigma^2_{\rm w} \left(\sum_{p=0}^{P-1}|h_p|^2 + \beta^2 |h_{\rm los}|^2\right)}
\end{equation}
as the SNR loss due to the additional interference terms.
The approximations of \eqref{eq:Y_uw} and \eqref{eq:W_uw} are due to the same reasons as the steps of \eqref{eq:Y_ofdm}.

\subsection{UW2 Frame}
In the case of the UW2 transmission, the radar samples the received signal as
\begin{equation}\label{eq:Y_uw2}
	\begin{split}
	&	\rev{\mathbf{Y}_{\rm uw2}[k,m] \approx \sum_{p=0}^{P-1}h_p' \mathbf{u}_{M,\tilde{\nu}_p}[m] (\mathbf{x}_{\rm uw2} \circledast \mathbf{g}_{N_{\rm cp}/2,\tilde{\tau}_p})[k]} \\ & \rev{+ \beta h_{\rm los}'\mathbf{u}_{M,\tilde{\nu}_{\rm los}}[m] (\mathbf{x}_{\rm uw2} \circledast \mathbf{g}_{N_{\rm cp}/2,\tilde{\tau}_{\rm los}})[k] \!+\! \mathbf{W}_{\rm uw2}[k,m]}
	\end{split}
\end{equation}	
for $0\leq k < N_{\rm cp}/2$ and $ 0\leq m < M$ such that $\mathbf{Y}_{\rm uw2} \in \mathbb{C}^{N_{\rm cp}/2 \times M}$.
$\mathbf{Y}_{\rm uw2}$ has an analogous structure to $\mathbf{Y}_{\rm ofdm}$ in \eqref{eq:Y_ofdm} by following the same steps.
\rev{Also, the noise sample of $\mathbf{W}_{\rm uw2}$ have the noise power of $\mathbf{W}_{\rm ofdm}$, $\sigma_{\rm w}^2$.}
Differently from the other schemes, for the UW2 signal, it is assumed that the normalized delay is within the interval $0\leq \tilde{\tau}_p < N_{\rm cp}/2$, which decreases the maximum delay in which the radar can detect targets.
Lastly, we note that the approximation in the second line of \eqref{eq:Y_uw2} is due to the same reasons as the steps of \eqref{eq:Y_ofdm}.

\section{Targets Delay-Doppler (DD) Estimation}\label{sec:delay_doppler_estimation}
\rev{This section starts with formulation of the DD MLE and \ac{CRB} as benchmark.
Then, to avoid the expensive computation of the MLE, we study a more practical solution for the multi-target DD estimation with LoS interference removal with reduced complexity that has two steps. 
First, the 2D channel is estimated in the DD domain with FFT processing for both target and LoS branches of the receiver depicted in Fig.~\ref{fig:system_model_high_level}.
Subsequently, the \emph{Fine Grid} algorithm is presented  that removes inter-target and LoS interference.}
%
%
\subsection{Maximum Likelihood Estimator (MLE)}\label{subsec:MLE}
%
\rev{Assuming perfect LoS removal by setting $\beta=0$, the MLE estimates DD pairs and channel gains of all reflections by minimizing the log-likelihood function 
\begin{equation}\label{eq:ml_estimator_theta}
\hat{\boldsymbol{\Theta}} =\min_{\boldsymbol{\Theta}} \sum_{k,m}\left|\mathbf{Y}[k,m] - \mathcal{Y}_{\boldsymbol{\Theta}}[k,m] \right|^2,
\end{equation}
where $\mathbf{Y}[k,m]$ can be OFDM, PS, UW1 or UW2 as described in Section \ref{sec:discrete_time_RX_signal} and $\mathcal{Y}_{\boldsymbol{\Theta}}[k,m]$ is the sampled noiseless received signal \eqref{eq:y_approx}  without LoS interference 
\begin{align}\label{eq:Ytheta}
& \mathcal{Y}_{\boldsymbol{\Theta}}[k,m] = \nonumber\\ & \sum_{p=0}^{P \!-\!1} |h_p| e^{j \angle h_p} e^{\!-j 2 \pi \tilde{\nu}_p \frac{I(k,m)}{M(N_{\rm cp}+K)}}\! \sum_{n=0}^{N_{\rm x}\!-\!1}\!\mathbf{x}[n]\Tilde{g}_{\rm RC}(I(k,m)\!- \! n \! -\! \tilde{\tau}_p),	
\end{align}
with the true parameters $\boldsymbol{\Theta} = [\boldsymbol{\theta}_0^{\rm T} \, \boldsymbol{\theta}_1^{\rm T} \, \cdots  \,\boldsymbol{\theta}_{P-1}^{\rm T}]^{\rm T} \in \mathcal{R}^{4P \times 1}$ that is the concatenation of the parameters per target $\boldsymbol{\theta}_p = [\tilde{\tau}_p \,\, \tilde{\nu}_p \,\, |h_p| \,\, \angle{h_p} ]^{\rm T}$.
Also, we note that the sampling indexes $I(k,m)$ in \eqref{eq:Ytheta} should be set accordingly for OFDM, PS, UW1 or UW2.}

\setcounter{mytempeqncnt}{\value{equation}}
\setcounter{equation}{32}
\begin{figure*}[b]
	\hrulefill
	\begin{equation}\label{eq:g_rc_d1}
		\tilde{g}'_{\rm RC}(t) = \frac{d \tilde{g}_{\rm RC}(t)}{dt}	= \left\{\begin{matrix} 
			\frac{1}{t}\frac{\cos\left(\alpha \pi t\right) \left(\cos(\pi t) - {\rm sinc}(t)\right)}{1 - (2\alpha t)^2} + \frac{8 \alpha^2 t \cos(\alpha \pi t) \rm{sinc}(\pi t)}{\left(1 - (2\alpha t)^2\right)^2} - \frac{\alpha \pi \sin(\alpha \pi t) \rm{sinc}(\pi t)}{1 - (2\alpha t)^2}, & t \neq 0, t \neq \pm 1/(2\alpha)
			\\ 0, & t=0
			\\ 1/2 \alpha \left(\pi \cos\left(\pi/(2\alpha)\right) - 3\alpha \sin\left(\pi/(2\alpha)\right)\right), & t = \pm 1/(2\alpha)
		\end{matrix}\right.
	\end{equation}	
\end{figure*}
\setcounter{equation}{\value{mytempeqncnt}}

\subsection{Cram{\'{e}}r-Rao Bound (CRB)}\label{subsec:CRB}
The CRB for the likelihood function defined in the argument of the ML estimator in \eqref{eq:ml_estimator_theta} is given in the following.
The $4P \times 4P$ Fisher information matrix is given by
\begin{equation}\label{eq:I}
	\begin{split}
		[\mathbf{I}(\boldsymbol{\Theta})]_{i,j} = 
		 \frac{2}{{\sigma}_{\rm w}^2}\sum_{m,k}{\cal{R}}\left( \frac{\partial}{\partial \boldsymbol{\Theta}_i}\mathcal{Y}_{\boldsymbol{\Theta}}[m,k] \left(\frac{\partial}{\partial \boldsymbol{\Theta}_j}\mathcal{Y}_{\boldsymbol{\Theta}}[m,k]\right)^\dagger\right).
	\end{split}
\end{equation}
The partial derivative concerning the delay $\tau_p$ of the $p$th target is given by
\begin{align}
	&	\rev{\frac{\partial \mathcal{Y}_{\boldsymbol{\Theta}}[m,k]}{\partial \tilde{\tau}_p} =}  \nonumber\\ & 	\rev{-|h_p| e^{j \angle h_p} e^{-j 2 \pi \tilde{\nu}_p \frac{I(k,m)}{N_{\rm x}}}	\sum_{n=0}^{N_{\rm x}-1}\mathbf{x}[n]\tilde{g}'_{\rm RC}(I(k,m)-n-\tilde{\tau}_p)	}
\nonumber \\ & 	\rev{+\!\sum_{\substack{q=0\\q\neq p}}^{P \!-\!1} |h_q| e^{\!j \angle h_q} e^{\!-j 2 \pi \tilde{\nu}_q \! \frac{I(k,m)}{M(N_{\rm cp}+K)}}\!\! \sum_{n=0}^{N_{\rm x}\!-\!1}\!\mathbf{x}[n]\Tilde{g}_{\rm RC}(I(k,\! m)\!- \! n \! -\! \tilde{\tau}_q),}
	\end{align}
%
%
where the derivative of the raised cosine filter is shown in \eqref{eq:g_rc_d1} \rev{at the bottom of the next page}, whose details are omitted due to lack of space and have been validated using the Wolfram Mathematica software. The values for $t = 0$ and $\pm 1/(2\alpha)$ are defined by taking the limit of the function in the first line of \eqref{eq:g_rc_d1}. 
The remaining partial derivatives with respect to $\nu_p$, $|h_p|$, and $\angle h_p$ are straightforward and are omitted due to the lack of space.
%
%
%
%
%
Finally, the CRB for the delay and Doppler parameters are computed as
\setcounter{equation}{25}
\begin{align}
&\rev{	{\rm CRB}_{\rm delay} = [\mathbf{I}(\boldsymbol{\Theta})^{-1}]_{1+pP,1+pP} } \nonumber \\ & \rev{{\rm CRB}_{\rm Doppler} = [\mathbf{I}(\boldsymbol{\Theta})^{-1}]_{2+pP,2+pP}.}
\end{align}
%
%
\subsection{FFT-based 2D Channel Estimation, Targets Branch}\label{subsec:2d_channel_estimation}
\subsubsection{OFDM} 
%
The channel $\hat{\mathbf{H}}_{\rm ofdm} \in \mathbb{C}^{M\times K}$ is estimated as
\begin{align}\label{eq:H_ofdm}
		\rev{\hat{\mathbf{H}}_{\rm ofdm}} & \rev{= \mathbf{F}_M^{\rm H}(\mathbf{F}_K^{\rm H}((\mathbf{F}_K\mathbf{Y}_{\rm ofdm}) \oslash \mathbf{D}_{\rm t}))^{\rm T}}
		\nonumber\\ & \rev{\approx  \sum_{p=0}^{P-1}h_p'\mathbf{v}_{M,\tilde{\nu}_p} \mathbf{g}_{K,\tilde{\tau}_p}^{\rm T} +  \mathbf{W}'_{\rm los} + \mathbf{W}'_{\rm ofdm}.}
\end{align}
A proof that \eqref{eq:H_ofdm} holds is provided in Appendix~\ref{ap_subsec:2D_channel_estimation} for completeness.
\rev{Also, we note because the OFDM targets and LoS streams, $s_{\rm t}(t)$ and $s_{\rm los}(t)$, have different data symbols, the LoS interference component $\mathbf{W}'_{\rm los}$ in \eqref{eq:H_ofdm} becomes convoluted.
Moreover, in this paper we do not evaluate the LoS interference removal for OFDM, thus we keep $\mathbf{W}'_{\rm los}$ without its explicit formulation for simplicity.}
The matrix $\mathbf{D}_{\rm t} = \left[\mathbf{d}_{{\rm t}_0} \,\,\, \mathbf{d}_{{\rm t}_1} \,\,\, \cdots \,\,\, \mathbf{d}_{{\rm t}_{M-1}}\right]	\in \mathbb{C}^{K \times M}$ stacks the QAM sub-blocks along rows.
The vector $\mathbf{v}_{M,\tilde{\nu}_l} \in \mathbb{C}^{M}$
\begin{equation}
	\mathbf{v}_{M,\tilde{\nu}} = \mathbf{F}_M^{\rm H}\mathbf{u}_{M,\tilde{\nu}}
\end{equation}
is simply the \ac{IDFT} of the complex exponential containing the Doppler phase $\rev{\mathbf{u}_{M,\tilde{\nu}}[m] = \exp(-j 2 \pi \tilde{\nu}m/M)}$.
Intuitively, one can easily note that for integer values of the normalized Doppler $\tilde{\nu} = 0, \pm 1, \pm 2, \cdots$,  the IDFT $\mathbf{v}_{M,\tilde{\nu}}[m] = \sqrt{M}$ for $m=\pm \tilde{\nu}$ and 0 otherwise\footnote{A proper FFT shift operation needs to be done to account for negative $\tilde{\nu}$.}, which allows an estimation of the Doppler shift.
Since the noise is divided by the QAM symbols in the frequency domain in \eqref{eq:H_ofdm}, its power is
%
%
$\tilde{\sigma}_{\rm w}^2 = \sigma_{\rm w}^2\mathbb{E}(1/(d d^{\rm H}))= \sigma_{\rm w}^2 L_{\rm ofdm}$ for all $(m,k)$, in which
\begin{equation}\label{eq:L_loss_ofdm}
	L_{\rm ofdm} = \frac{1}{|\mathcal{S}|}\sum_{{\rm d} \in \mathcal{S}}  \frac{1}{{\rm d} {\rm d}^{\rm H}},
\end{equation}
being $\mathcal{S}$ the QAM set.
Since $L_{\rm ofdm} \geq 1$, with $L_{\rm ofdm} = 1$ when ${\rm d} {\rm d}^{\rm H} = 1 \forall {\rm d}$, $L_{\rm ofdm}$ can be interpreted as the SNR loss in relation to the constant amplitude pilot signaling.

Lastly, it is worth noting that the operation of \eqref{eq:H_ofdm} is similar to the \ac{SFFT} carried out in DD processing in \ac{OTFS}. 
But here we have derived it with equalization in the frequency domain outside the scope of OTFS.
This enables a common framework for the UW frames of Subsection~\ref{subsec:tx_signals_frames}.

\subsubsection{PS}
The estimator for the PS frame is similar to the CP-OFDM case, but the input is $\mathbf{Y}_{\rm ps} \in \mathbb{C}^{K \times M_{\rm p}}$ in \eqref{eq:Y_prs} so that the estimated channel is
\begin{align}
		&\hat{\mathbf{H}}_{\rm ps}  =\mathbf{F}_{M_{\rm p}}^{\rm H}(\mathbf{F}_K^{\rm H}((\mathbf{F}_K\mathbf{Y}_{\rm ps}) \oslash \mathbf{P}))^{\rm T}
		\nonumber\\ & \rev{\approx   \sum_{p=0}^{P-1}h_p'\mathbf{v}_{M_{\rm p},\tilde{\nu}_p} \mathbf{g}_{K,\tilde{\tau}_p}^{\rm T} + \beta h_{\rm los}'\mathbf{v}_{M_{\rm p},\tilde{\nu}_{\rm los}} \mathbf{g}_{K,\tilde{\tau}_{\rm los}}^{\rm T} + \mathbf{W}_{\rm ps}',}
\end{align}	
with size $\hat{\mathbf{H}}_{\rm ps} \in \mathbb{C}^{M_{\rm p}\times K}$. Analogously to $\mathbf{D}_{\rm t}$ in \eqref{eq:H_ofdm}, $\mathbf{P} = \left[\mathbf{p}_0 \,\,\, \mathbf{p}_1 \,\,\, \cdots \,\,\, \mathbf{p}_{M_{\rm p}-1} \right]	\in \mathbb{C}^{K \times M_{\rm p}}$ stacks the PS symbols along rows.
And $\mathbf{v}_{M_{\rm p},\tilde{\nu}} = \mathbf{F}_{M_{\rm p}}^{\rm H}\mathbf{u}_{M_{\rm p},\tilde{\nu}}$ has size $M_{\rm p}$.
The power of $\mathbf{W}_{\rm ps}'$ is equal to $\sigma_{\rm w}^2$ because $\mathbf{p}_m$ has constant amplitude.
\rev{Lastly, we note that the LoS interference has a convenient form because the pilot symbols are the same for both targets and LoS streams.}

\subsubsection{UW1} 
Following the same approach as the OFDM signaling, but with $\mathbf{Y}_{\rm uw1} \in \mathbb{C}^{N_{\rm cp} \times M}$ as input, the channel estimation for the UW1 signal is
\begin{align}\label{eq:H_uw}
		& \hat{\mathbf{H}}_{\rm uw1} = \mathbf{F}_{M}^{\rm H}(\mathbf{F}_{N_{\rm cp}}^{\rm H}((\mathbf{F}_{N_{\rm cp}}\mathbf{Y}_{\rm uw1}) \oslash \mathbf{X}_{\rm uw1}))^{\rm T}
	\nonumber	\\ & \rev{\approx \sum_{p=0}^{P-1} h_p'\mathbf{v}_{M,\tilde{\nu}_p} \mathbf{g}_{N_{\rm cp},\tilde{\tau}_p}^{\rm T}\! + \! \beta h_{\rm los}'\mathbf{v}_{M,\tilde{\nu}_{\rm los}} \mathbf{g}_{N_{\rm cp},\tilde{\tau}_{\rm los}}^{\rm T} + \! \mathbf{W}_{\rm uw1}',}
\end{align}	
with size $\hat{\mathbf{H}}_{\rm uw1} \in \mathbb{C}^{M\times N_{\rm cp}}$, where $\mathbf{X}_{\rm uw1} = \mathbf{F}_{N_{\rm cp}}\left[\mathbf{x}_{\rm uw1} \,\,\, \mathbf{x}_{\rm uw1} \,\,\, \cdots \,\,\, \mathbf{x}_{\rm uw1}\right]	\in \mathbb{C}^{N_{\rm cp} \times M}$ stacks the UW1 of size $N_{\rm cp}$ in frequency-domain along rows, with the time-domain raised cosine filter of size $N_{\rm cp}$.
\rev{Since the noise plus interference term $\mathbf{W}_{\rm uw1}$ in \eqref{eq:W_uw} is independent of the UW1, and the UW1 has constant amplitude in the frequency domain, it is easy to verify that the resulting noise plus interference $\mathbf{W}_{\rm uw1}'$  in \eqref{eq:H_uw} has the same power as $\mathbf{W}_{\rm uw1}$, which is approximated by \eqref{eq:noise_power_uw} with the SNR loss $L_{\rm uw1}$ given in \eqref{eq:L_loss_uw}.}
%
%

\subsubsection{UW2} 
Let the UW2 input be $\mathbf{Y}_{\rm uw2} \in \mathbb{C}^{N_{\rm cp}/2 \times M}$, the channel estimation is expressed as
\begin{align}
		& \hat{\mathbf{H}}_{\rm uw2}  = \mathbf{F}_{M}^{\rm H}(\mathbf{F}_{N_{\rm cp}/2}^{\rm H}((\mathbf{F}_{N_{\rm cp}/2}\mathbf{Y}_{\rm uw2}) \oslash \mathbf{X}_{\rm uw2}))^{\rm T}
				\nonumber\\ & \rev{\approx \sum_{p=0}^{P-1} h_p'  \mathbf{v}_{M,\tilde{\nu}_p} \mathbf{g}_{N_{\rm cp}/2,\tilde{\tau}_p}^{\rm T} +  \beta h_{\rm los}'  \mathbf{v}_{M,\tilde{\nu}_{\rm los}} \mathbf{g}_{N_{\rm cp}/2,\tilde{\tau}_{\rm los}}^{\rm T} + \mathbf{W}_{\rm uw2}',}
\end{align}	
with size $\hat{\mathbf{H}}_{\rm uw2} \in \mathbb{C}^{M\times N_{\rm cp}/2}$, where 	$\mathbf{X}_{\rm uw2} = \mathbf{F}_{N_{\rm cp}/2}\left[\mathbf{x}_{\rm uw2} \,\,\, \mathbf{x}_{\rm uw2} \,\,\, \cdots \,\,\, \mathbf{x}_{\rm uw2}\right]	\in \mathbb{C}^{N_{\rm cp}/2 \times M}$ stacks the UW2 signals of size $N_{\rm cp}/2$ in frequency-domain along rows.
In this case, the time-domain raised cosine filter has a size of $N_{\rm cp}/2$.
The power of $\mathbf{W}_{\rm uw2}'$ is equal to $\sigma_{\rm w}^2$ because $\mathbf{\mathbf{x}}_{\rm uw2}$ has constant amplitude.

\subsection{\rev{LoS Channel Estimation}}\label{subsec:2d_channel_estimation_los} 
\rev{An analogous formulation for the LoS branch in Fig.~\ref{fig:system_model_high_level} is straightforward to obtain by following the same steps of the previous subsection.
As an example, the estimated LoS for the UW2 frame is}
\setcounter{equation}{33}
\begin{align}
	\rev{\hat{\mathbf{H}}_{\rm los_{uw2}} \approx \sum_{p=0}^{P-1} h_{\rm los}'  \mathbf{v}_{M,\tilde{\nu}_{\rm los}} \mathbf{g}_{N_{\rm cp}/2,\tilde{\tau}_{\rm los}}^{\rm T} + \mathbf{W}_{\rm los_{uw2}}'.}
\end{align}	
\rev{The other frames can be obtained by an analogous formulation.
	The interference from the target reflections into the LoS branch is incorporated in $\mathbf{W}_{\rm los_{uw2}}$ for simplicity, and it is assumed to be negligible due to the much higher gain of the LoS path in comparison to the paths associated to the targets.}
%
\subsection{Fine Grid DD Estimation}\label{subsec:fine_grid_estimation}
%
Firstly, we define the DD signature matrix for the channel estimation schemes of Subsection~\ref{subsec:2d_channel_estimation} as
\begin{equation}\label{eq:psi}
	\mathbf{\Psi}_{\tilde{\nu},\tilde{\tau}} =  \left\{\begin{matrix}
		\mathbf{v}_{M,\tilde{\nu}} \mathbf{g}_{K,\tilde{\tau}}^{\rm T}, & \text{OFDM}	
		\\ 	\mathbf{v}_{M_{\rm p},\tilde{\nu}} \mathbf{g}_{K,\tilde{\tau}}^{\rm T}, & \text{PS}	
		\\ \mathbf{v}_{M,\tilde{\nu}} \mathbf{g}_{N_{\rm cp},\tilde{\tau}}^{\rm T}, & \text{UW1}	
		\\ 	\mathbf{v}_{M,\tilde{\nu}} \mathbf{g}_{N_{\rm cp}/2,\tilde{\tau}}^{\rm T}, & \text{UW2}		
	\end{matrix}\right..
\end{equation}
\rev{In the following we provide an iterative multi-user estimator similar to \cite{Gaudio} where the estimates are refined by removing the interference from the targets.
When the DD estimates for the LoS and the $q$th target are available $\{ \hat{\tilde{\nu}}_q,\hat{\tilde{\tau}}_q\}$, their interference can be removed as}
\begin{equation}\label{eq:Hhat_p}
\rev{\hat{\mathbf{H}}_p = \hat{\mathbf{H}} - \sum_{\substack{q=0 \\ q\neq p}}^{P-1} \hat{h}_q'  \mathbf{\Psi}_{\hat{\tilde{\nu}}_q,\hat{\tilde{\tau}}_q} - \widehat{(\beta h_{\rm los}')}  \mathbf{\Psi}_{\hat{\tilde{\nu}}_{\rm los},\hat{\tilde{\tau}}_{\rm los}},}
\end{equation}
\rev{which is valid for PS, UW1 and UW2 and not for OFDM since its LoS interference has a different structure.
Then, using the 2D channel estimation as the observation, the estimator for the $p$th DD target is given as}
\begin{equation}\label{eq:dd_estimation}
	\rev{\{ \hat{\tilde{\nu}}_p,\hat{\tilde{\tau}}_p\} = \max_{\substack{\tilde{\nu},\tilde{\tau}}} \frac{\left|\sum_{m,k}\hat{\mathbf{H}}_p[m,k] \mathbf{\Psi}_{\tilde{\nu},\tilde{\tau}}[m,k]^{\dagger}\right|^2}{\sum_{m,k}| \mathbf{\Psi}_{\tilde{\nu},\tilde{\tau}}[m,k]|^2},}
\end{equation}
\rev{and the channel gain is estimated as}
\begin{equation}\label{eq:hhat_p}
\rev{\hat{h}_p' =  \frac{\hat{\mathbf{H}}_p[{\rm round}(\hat{\tilde{\nu}}_p),{\rm round}(\hat{\tilde{\tau}}_p)]}{\mathbf{\Psi}_{\hat{\tilde{\nu}}_p,\hat{\tilde{\tau}}_p}[{\rm round}(\hat{\tilde{\nu}}_p),{\rm round}(\hat{\tilde{\tau}}_p)]},}
\end{equation}
\rev{where the channel gain and phase of the $p$th target is estimated to allow proper removal in  \eqref{eq:Hhat_p}. 
Since the channel estimate $\hat{\mathbf{H}}_p$ is discrete, the DD estimated $	\{ \hat{\tilde{\nu}}_p,\hat{\tilde{\tau}}_p\}$ are approximated to the nearest integer to take the peak sample and avoid taking noisy samples in the channel gain estimation.
Likewise, the LoS DD is estimated as}
\begin{equation}\label{eq:dd_estimation_los}
\rev{	\{ \hat{\tilde{\nu}}_{\rm los},\hat{\tilde{\tau}}_{\rm los}\} = \max_{\substack{\tilde{\nu},\tilde{\tau}}} \frac{\left|\sum_{m,k}\hat{\mathbf{H}}_{\rm los}[m,k] \mathbf{\Psi}_{\tilde{\nu},\tilde{\tau}}[m,k]^{\dagger}\right|^2}{\sum_{m,k}| \mathbf{\Psi}_{\tilde{\nu},\tilde{\tau}}[m,k]|^2},}
\end{equation}
\rev{and the channel gain is estimated as}
\begin{equation}\label{eq:h_los_hat}
\rev{	\widehat{(\beta h_{\rm los}')} =  \frac{\hat{\mathbf{H}}[{\rm round}(\hat{\tilde{\nu}}_{\rm los}),{\rm round}(\hat{\tilde{\tau}}_{\rm los})]}{\mathbf{\Psi}_{\hat{\tilde{\nu}}_{\rm los},\hat{\tilde{\tau}}_{\rm los}}[{\rm round}(\hat{\tilde{\nu}}_{\rm los}),{\rm round}(\hat{\tilde{\tau}}_{\rm los})]}.}
\end{equation}
\rev{Notice that the LoS branch is used to estimate the LoS DD $\{ \hat{\tilde{\nu}}_{\rm los},\hat{\tilde{\tau}}_{\rm los}\} $ in \eqref{eq:dd_estimation_los}.
Then, the interference gain is computed in \eqref{eq:h_los_hat} using the estimated channel in the targets branch so that it can be removed accordingly in \eqref{eq:Hhat_p}.
Furthermore, we highlight that the computation of \eqref{eq:dd_estimation} and \eqref{eq:dd_estimation_los} is implemented via a fine grid search whose complexity increases linearly with the desired grid resolution, thereby yielding the name \emph{Fine Grid} estimator. 
The pseudocode for the multi-target DD estimation is presented in Algorithm 1, where the last step, line 13, is responsible for the offset cancellation in the delay and Doppler domains.}

\revv{Lastly, we highlight that the presented algorithm relies on the assumption of a single path per target of \eqref{eq:r}. If each target produces multiple paths with a small delay and Doppler spread, the performance tends to degrade.
	This analysis is left for future work.}
%
%
\begin{algorithm}[t!]
	\caption{\rev{Multi-Target Delay-Doppler Estimation}}
	\rev{\begin{algorithmic}[1]
		\State Estimate $\hat{\mathbf{H}}$ according to Subsection \ref{subsec:2d_channel_estimation} 
		\State Estimate $\hat{\mathbf{H}}_{\rm los}$ according to Subsection \ref{subsec:2d_channel_estimation_los}
		\State Estimate $\{ \hat{\tilde{\nu}}_{\rm los},\hat{\tilde{\tau}}_{\rm los}\}$ with \eqref{eq:dd_estimation_los}
		\State Estimate $\widehat{(\beta h_{\rm los}')}$ with \eqref{eq:h_los_hat}
		\State $\hat{h}_p' \gets 0, \hat{\tilde{\nu}}_p \gets 0, \hat{\tilde{\tau}}_p \gets 0 \, \forall p$
		\For {$n_{\rm i} =1,\ldots, N_{\rm iterations}$}
		\For {$p =1,\ldots,P$}
			\State Estimate $\hat{\mathbf{H}}_p$ with \eqref{eq:Hhat_p}
			\State Estimate $\{ \hat{\tilde{\nu}}_p,\hat{\tilde{\tau}}_p\}$ with \eqref{eq:dd_estimation}
			\State Estimate $\hat{h}_p'$ with \eqref{eq:hhat_p}
		\EndFor
		\EndFor
		\State Compute $\hat{\tilde{\tau}}_{{\Delta}_p} \gets \hat{\tilde{\tau}}_p - \hat{\tilde{\tau}}_{\rm los}$ and $\hat{\tilde{\nu}}_{{\Delta}_p} \gets \hat{\tilde{\nu}}_p - \hat{\tilde{\nu}}_{\rm los} \, \forall p$
	\end{algorithmic}}
\end{algorithm}
%

\section{Outlier Probability of Integer Grid Estimator}\label{sec:outlier_probability}
%
%

\subsection{\rev{Integer Grid Estimator}}\label{subsec:integer_grid}
In practical systems, the maximum likelihood or fine grid estimators of \eqref{eq:ml_estimator_theta} and \eqref{eq:dd_estimation} might not be available for several reasons. 
For example, its complexity can be prohibitive, and the knowledge of the filter responses is not perfect making its implementation more difficult and less effective.
Thus, a more practical solution is the estimation of the \emph{Integer Grid} as
\begin{equation}\label{eq:mk_hat}
	{\{ \hat{m},\hat{k}\} = \max_{m,k} |\hat{\mathbf{H}}[m,k]|^2.}
\end{equation}
In this work, we focus on the outlier probability for the single-target case for simplicity, thus we keep the formulation \eqref{eq:mk_hat} for this case.
A generalization for the multiple targets case is possible by taking into account the mutual interference that the targets impose in each other.
%
%
%
The sets $\mathcal{M}$ and $\mathcal{K}$ contain possible Doppler and delay indexes depending on the transmitted frame
\begin{equation}\label{eq:K_set}
	{\mathcal{K} = \left\{\begin{matrix}
			\{0,1,\cdots,N_{\rm cp}-1\}, & \text{OFDM, PS \& \rev{UW1}}
			\\ \{0,1,\cdots,N_{\rm cp}/2-1\}, & \text{UW2}
		\end{matrix}\right.}
\end{equation}
and
\begin{equation}\label{eq:M_set}
{	\mathcal{M} = \left\{\begin{matrix}
			\{0,1,\cdots,M-1\}, & \text{OFDM, UW1 \& UW2}
			\\ \{0,1,\cdots,M_{\rm p}\}, & \text{PS}
		\end{matrix}\right..}
\end{equation}
%

%
%
\subsection{Outlier Probability}
\rev{By decoupling the DD into its integer and fractional components, a convenient metric to assess the performance of \eqref{eq:mk_hat} is the outlier probability, which measures the probability that the integer DD estimation is wrong.}
Defining the DD by its integer and fractional components as $\tilde{\tau} = \tau_{\rm I} + \tau_{\rm F}$	and $\tilde{\nu} = \nu_{\rm I} + \nu_{\rm F}$,
%
%
the 2D indexes that are mapped to the integer DD are $k_0 =  \tau_{\rm I} $ for the time index, and $m_0 = \nu_{\rm I}$ if $\nu_{\rm I} < M/2$ and $m_0 = M-\nu_{\rm I}$ if $\nu_{\rm I} \geq M/2$ for the Doppler index. Notice that a simple transformation is needed to account for negative Doppler.

\begin{figure*}[b]
	\hrulefill
	\setcounter{mytempeqncnt}{\value{equation}}
	\setcounter{equation}{51}
	\begin{equation}\label{eq:pdf_X_geq_Y}
			{\rm Pr}(X>Y)  = {\rm Pr}(X/Y>1) = \frac{1}{2}\exp \left( -\frac{v_{\rm x}^2 + v_{\rm y}^2}{2\sigma^2}\right)\sum_{k=0}^{\infty}\frac{1}{k!}\left(\frac{v_{\rm x}^2}{2\sigma^2}\right)^{k} \sum_{j=0}^{\infty} \frac{1}{j!}\left(\frac{v_{\rm y}^2}{4}\right)^{j} {_2F_1}\left(-j,-j;k+1;\frac{v_{\rm x}^2}{v_{\rm y}^2}\right)
		\end{equation}
		\setcounter{equation}{\value{mytempeqncnt}}
	\end{figure*}
	
The delay outlier probability for the delay estimation can be approximated by
\begin{align}\label{eq:P_delay}
	{P}_{{\rm delay}} & \approx {\rm Pr}\left(\bigcup_{\substack{k \in \mathcal{K} \\ k \neq k_0}} 	\left\{|\hat{\mathbf{H}}[m_0,k]| > |\hat{\mathbf{H}}[m_0,k_0]|\right\}\right)
	\nonumber\\ & \lesssim \sum_{\substack{k\in \mathcal{K} \\ k \neq k_0}} {\rm Pr}(|\hat{\mathbf{H}}[m_0,k]| > |\hat{\mathbf{H}}[m_0,k_0]|).	
\end{align}
The first line of \eqref{eq:P_delay} computes the delay outlier probability assuming that the Doppler index is correctly estimated, where $\mathcal{K}$ is given in \eqref{eq:K_set}.
\rev{In particular, the event $|\hat{\mathbf{H}}[m_0,k]| > |\hat{\mathbf{H}}[m_0,k_0]|$ for an arbitrary $k \neq k_0$ indicates a situation where the estimator \eqref{eq:mk_hat} would choose the wrong index $k$ instead of $k_0$. Since this can happen for multiple $k\neq k_0$, the outlier probability formulation takes the union of the events for all $k\neq k_0$.} 
The second line is the well-known union bound or Boole's inequality, which counts the events' intersections more than once.
For the Doppler estimation, the analogous formulation is
\begin{equation}\label{eq:P_doppler}
	\begin{split}
		P_{{\rm Doppler}} \lesssim \sum_{\substack{m\in \mathcal{M} \\ m \neq m_0}} {\rm Pr}(|\hat{\mathbf{H}}[m,k_0]| > |\hat{\mathbf{H}}[m_0,k_0]|),
	\end{split}
\end{equation}
where the delay index is assumed to be correctly estimated and $\mathcal{M}$ is given in \eqref{eq:M_set}.

We note the absolute value of the channel estimate $\hat{\mathbf{H}}[m,k]|$ in \eqref{eq:P_delay} and \eqref{eq:P_doppler} follows a Rician distribution. 
Let $R\sim {\rm Rice}(v,\sigma)$ be a Rician random variables with parameters $(v,\sigma)$ and \ac{PDF} given by
\begin{equation}\label{eq:pdf_rice}
	f_R(r;v,\sigma) = \frac{r}{\sigma^2}	\exp\left( \frac{-(r+v)^2}{2 \sigma^2} \right) I_0\left(\frac{r v}{\sigma^2}\right),
\end{equation} 
where $I_0(\cdot)$ is the modified Bessel function of the first kind with order zero.
We have $|\hat{\mathbf{H}}[m,k]| \sim {\rm Rice}(v_{m,k},\sigma)$ with
\begin{equation}\label{eq:v}
	v_{m,k} \approx |h \mathbf{\Psi}_{\tilde{\nu},\tilde{\tau}}[m,k]|,
\end{equation}	
and 
\begin{equation}\label{eq:sigma}
	\sigma^2 = \left\{\begin{matrix}
		1/2\sigma_{\rm w}^2L_{\rm ofdm}, & \text{CP-OFDM}
		\\ 1/2\sigma_{\rm w}^2, & \rev{\text{PS}}
		\\ 1/2\sigma_{\rm w}^2L_{\rm uw1}, & \text{UW1}
		\\ 1/2\sigma_{\rm w}^2, & \text{UW2}		
	\end{matrix}\right.,
\end{equation}
where $\mathbf{\Psi}_{\tilde{\nu},\tilde{\tau}}$ and is given by \eqref{eq:psi}.
%
%
Thus, the problem of computing \eqref{eq:P_delay} and \eqref{eq:P_doppler} consists of computing the probability that a Rician \ac{RV} is greater than another Rician RV, which is presented in the following.
%
%

\subsection{Probability that a Rician Variable is Greater than Another
}\label{ap:rice_greater_rice} 
%
%
Let $X\sim {\rm Rice}(v_{\rm x},\sigma)$ and $Y\sim {\rm Rice}(v_{\rm y},\sigma)$ be independent Rician random variables with PDF given by \eqref{eq:pdf_rice}.
We are interested in computing the probability that $X$ is greater than $Y$, which can be written using the \ac{CDF} of the ratio $X/Y$ by writing ${\rm Pr}(X>Y) = {\rm Pr}(X/Y>1)$.
The CDF of the ratio of independent Rician random variables has been derived in \cite{Kavous}, and is given in \eqref{eq:pdf_X_geq_Y} \rev{at the bottom of this page}, where $_2F_1(\cdot,\cdot;\cdot;\cdot)$ is Gauss hypergeometric function. 

Unfortunately, the numerical computation of \eqref{eq:pdf_X_geq_Y} is not feasible in general due to the summations to infinity.
To solve this issue, we provide two alternative solutions to compute \eqref{eq:pdf_X_geq_Y} approximately with affordable complexity.
	\subsubsection{Empirical Approximation (low and medium SNR)}
	For simplicity, we first normalize $X' = X/\sigma \sim {\rm Rice}(v_{\rm x}',1)$  and $Y' = Y/\sigma \sim {\rm Rice}(v_{\rm x}',1)$, with $v_{\rm x}' = v_{\rm x}/\sigma$ and $v_{\rm y}' = v_{\rm y}/\sigma$.
	Clearly, ${\rm Pr}(X>Y) = {\rm Pr}(X'>Y')$.
	
	By fixing $v_{\rm x}'$ and computing ${\rm Pr}(X>Y)$ for different values of $v_{\rm y}'$ with $v_{\rm y}'>v_{\rm x}'$, we realize that the resulting function behaves approximately as the Gaussian function, whose parameters depend on $v_{\rm x}$.
	Our approach is to compute the Gaussian parameters for some specific values of $v_{\rm x}'$ using \eqref{eq:pdf_X_geq_Y} and then fit these values to a function such that a general expression is derived.
	The result is given below
	\begin{align}\label{eq:approx1}
			& {\rm Pr}(X>Y) \approx \nonumber\\ & 
			\left\{\begin{matrix}
					\frac{1}{2}\exp  \left( \frac{(v_{\rm x}' - f_2(v_{\rm x}'))^2-(v_{\rm y}' - f_2(v_{\rm x}'))^2 }{2f_1(v_{\rm x}')}\right), & v_{\rm x}<v_{\rm y}\\
				1- \frac{1}{2}\exp  \left( \frac{(v_{\rm x}' - f_2(v_{\rm x}'))^2-(v_{\rm y}' - f_2(v_{\rm x}'))^2 }{2f_1(v_{\rm x}')}\right), & v_{\rm x} \geq v_{\rm y}
			\end{matrix}\right.
	\end{align}
	where 
	\begin{equation}
		\begin{split}
			&f_1(v) = 2+\\ & 0.6616 \left(1 - \exp\left(-1.909v^{1.3838}  \right)  \left(1 + 1.909v^{1.3838} \right)  \right)
		\end{split}
	\end{equation}
	\begin{equation}
		f_2(v) = 1.4899  \exp\left(-\frac{1}{0.39899 v^{0.89899}  }\right)  v^{0.89899}.
	\end{equation}
Basically, \eqref{eq:approx1} has the Gaussian function format over $v_{\rm y}'$.
	This function has been normalized to satisfy ${\rm Pr}(X>Y)=1/2$ if $v_{\rm y}' = v_{\rm x}'$.
	The functions $f_1(v)$ and $f_2(v)$ are the fitting functions having $v_{\rm x}'$ as input, which were found numerically.
	This solution is a very good approximation for $0 \leq v_{\rm x}' < 30$.
	For $v_{\rm x}'>30$, the functions $f_1(v)$ and $f_2(v)$ do not approximate well the Gaussian function \eqref{eq:approx1}.
	By realizing that high values of $v_{\rm x}'$ imply high SNR, another approximation is possible in this regime which is studied in the following.

	\subsubsection{Gaussian Approximation (high SNR)}
	Again we consider ${\rm Pr}(X>Y) = {\rm Pr}(X/Y>1)$. 
	Next, we note that for high values of $v_{\rm x}/(2\sigma^2)$ and $v_{\rm y}/(2\sigma^2)$, $X$ and $Y$ approximate to Gaussian with low coefficients of variation $\delta_{\rm x} = \sqrt{\mathbb{V}(X)}/\mathbb{E}(X)$ and $\delta_{\rm y} = \sqrt{\mathbb{V}(Y)}/\mathbb{E}(Y)$, which is a condition for approximating the ratio $X/Y$ to Gaussian as shown in \cite{frances}.	
	Then, by letting $Z = X/Y$ be approximated to $\mathcal{N}(\mu_{\rm z},\sigma_{\rm z}^2)$, its  moments can be approximated as
	\setcounter{equation}{52}
	\begin{equation}\label{eq:mu_z}
		\mu_{\rm z}  = \mathbb{E}(X/Y) \stackrel{(a)}{\approx} \mathbb{E}(X)/\mathbb{E}(Y) \stackrel{(b)}{\approx} v_{\rm x}/v_{\rm y}
	\end{equation}	
	and
	\begin{align}\label{eq:sigma2_z}
			\sigma_{\rm z}^2 & = \mathbb{V}(X/Y) 
			\nonumber \stackrel{(a)}{\approx} \left(\frac{\mathbb{E}(X)^2}{\mathbb{E}(Y)^2}\right) \left( \frac{\mathbb{V}(X)^2}{\mathbb{E}(X)^2}+\frac{\mathbb{V}(Y)^2}{\mathbb{E}(Y)^2}\right)
		\nonumber	\\ &\stackrel{(c)}{\approx} v_{\rm x}^2/v_{\rm y}^2(2\sigma^2/v_{\rm x}^2 + 2\sigma^2/v_{\rm y}^2).
	\end{align}
	The approximations $(a)$ in \eqref{eq:mu_z} and \eqref{eq:sigma2_z} use the method of \cite[Example 5.5.27]{casella} by expanding the parametric function $g(X,Y) = X/Y$ around the mean of $X$ and $Y$ using the first-order Taylor series for both variables. 
	The approximation $(b)$ in \eqref{eq:mu_z} is attained due to $\mathbb{E}(X), \mathbb{E}(Y) \rightarrow v_{\rm x},v_{\rm y} \text{ as } v_{\rm x},v_{\rm y} \rightarrow \infty$, which becomes more accurate as $v_{\rm x}$ and $v_{\rm y}$ increases.
	The approximation $(c)$ in \eqref{eq:sigma2_z} is attained due $\mathbb{V}(X), \mathbb{V}(Y) \rightarrow 1 \text{ as } v_{\rm x},v_{\rm y} \rightarrow \infty$, which also becomes more accurate as $v_{\rm x}$ and $v_{\rm y}$ increases.
	Finally, we have
	\begin{equation}
		{\rm Pr}(X>Y) = {\rm Pr}(X/Y>1) \approx Q\left( \frac{1-\mu_{\rm z}}{\sigma_{\rm z}}\right),
	\end{equation}
    where $ Q(x) = \frac{1}{2} \int_{x}^{\infty} e^{-\frac{t^2}{2}} \, dt$.
	This approximation is used for $v_{\rm x}'\geq 30$.
\section{Frame Design Overview}\label{sec:frame_design_overview}
In this section, we provide a comparative overview of the frames described in Subsection~\ref{subsec:tx_signals_frames}.
The parameters analyzed in this section are described in Table~\ref{tab:frame_design_overview}.

The idea of using PS, UW1, and UW2 processing is to avoid the need for the passive radar unit to know the transmitted data $\mathbf{d}$, to facilitate the deployment.
Another relevant aspect is the loss in data rate when dedicated pilot symbols replace the OFDM sub-blocks, which happens in the PS frame type.
Since the PS frame spends $M_{\rm p}$ out of $M$ for the pilots, its data rate loss is given by $M_{\rm p}/M$.
The above considerations are indicated in Table~\ref{tab:frame_design_overview} and illustrate well the motivation behind considering the UW1 and UW2 frames, which simultaneously avoid the radar's need to know the data, and do not incur data rate loss for communications.

The remaining quantitative parameters are analyzed in the following subsections.
\begin{table*}[t!]
	\centering
	\footnotesize
	\caption{Comparison of the frames of Subsection \ref{subsec:tx_signals_frames}. Integer DD resolution, $\Delta_\tau = T_{\rm s}$, and $\Delta_\nu = B/N_{\rm x}$.}
	\vspace{-3mm}	
	\begin{tabular}{c|c|c|c|c|c|c|c}\toprule
		Frame 	   & Knowledge & Data Rate 	& Proc. & Complexity  & Max.  & Maximum & SNR \\ 
		Design     & of Data, $\mathbf{d}$ & Loss  & Gain &      & Delay  & Abs. Doppler & Loss \\ \midrule
		OFDM    & yes & 0 & $MK$ &  $MK(\log_2 MK^2 \!+\!1\!+\! \rev{N_{\rm fg}})$ &  $(N_{\rm cp}-1)\Delta_\tau$  & $(M/2-1)\Delta_\nu $ & \eqref{eq:L_loss_ofdm} \& Table~\ref{tab:snr_loss_ofdm} \\
		PS        & no & $M_{\rm p}/M$ & $M_{\rm p}K$ & $M_{\rm p}K (\log_2 M_{\rm p}K^2\!+\! 1 \!+\! \rev{N_{\rm fg}})$  & $(N_{\rm cp}-1)\Delta_\tau$ 	& $(M_{\rm p}/2-1)\Delta_\nu $ & no loss \\ 
		UW1    & no & 0 & $MN_{\rm cp}$  & $M N_{\rm cp} (\log_2 M N_{\rm cp}^2\!+\!1\!+\! \rev{N_{\rm fg}})$  & $(N_{\rm cp}-1)\Delta_\tau$  & $(M/2-1)\Delta_\nu $ &  \eqref{eq:L_loss_uw}\\ 
		UW2   & no & 0 & $M\frac{N_{\rm cp}}{2}$ & $M \frac{N_{\rm cp}}{2} (\log_2 M (\frac{N_{\rm cp}}{2})^2\!+\!1\!+\! \rev{N_{\rm fg}})$  & $(\frac{N_{\rm cp}}{2}-1)\Delta_\tau$  & $(M/2-1)\Delta_\nu $ & no loss \\ \bottomrule		\end{tabular}\label{tab:frame_design_overview}
		\vspace{-0.2cm}
\end{table*}

\subsection{Processing Gain}\label{subsection:processing_gain}
The processing gain is a typical radar parameter related to the gain in SNR due to coherent processing, which is proportional to the number of time-domain samples processed by the radar receiver.
For the passive radar schemes considered in this paper, the processing gain is exactly the number of samples per sub-block times the number of sub-blocks.
Then, the processing gain for each frame scheme is given by $G_{\rm ofdm} = MK$, $G_{{\rm ps}} = M_{\rm p}K$, $G_{\rm uw1} = MN_{\rm cp}$ and $G_{\rm uw2} = MN_{\rm cp}/2$.
\subsection{Delay-Doppler Resolution}
\revv{The \emph{Integer Grid} delay-Doppler resolution is the same for all frames, namely, $\Delta_\tau = T_{\rm s} = 1/B$ for time and $\Delta_\nu = B/N_{\rm x}$ for Doppler, and is valid for single and multiple targets.}
The time resolution is simply the sampling interval. 
For the Doppler resolution, it is the subcarrier spacing when the whole bandwidth is divided by the total number of transmitted samples $N_{\rm x}=M(N_{\rm cp}+K)$, which is the same for all frames.
Note that what differentiates the frames of Subsection \ref{subsec:2d_channel_estimation} is the samples in which the radar receiver processes, while the transmitted frames have all the same format in terms of size and number of sub-blocks.

The resolution for the \emph{Fine Grid} estimator is simply the \emph{Integer Grid} resolution divided by the number of times the \emph{Integer Grid} is partitioned.

\subsection{Maximum Unambiguous Delay-Doppler}\label{subsec:maximum_unambiguous_range}
The maximum delay-Doppler is the resolution multiplied by the maximum index of the \emph{Integer Grid}.
For the delay, these numbers are $\tau^{\rm max}_{\rm ofdm} = \tau^{\rm max}_{\rm ps} = \tau^{\rm max}_{\rm uw1} = (N_{\rm cp}-1)\Delta_\tau$, ${\tau^{\rm max}_{\rm uw2} = (N_{\rm cp}/2-1)\Delta_\tau}$.
The respective maximum absolute Doppler are $\nu^{\rm max}_{\rm ofdm} = \nu^{\rm max}_{\rm uw1} = \nu^{\rm max}_{\rm uw2} = (M/2-1) \Delta_\nu$, $\nu^{\rm max}_{\rm PS} = (M_{\rm p}/2-1)\Delta_\nu$.

\rev{We observe that the PS frame has reduced maximum Doppler because its processed sub-blocks have a larger time separation, which causes an effect of increasing the channel sampling interval by $M/M_{\rm p}$ times in relation to the other frames.
Also, the UW2 frame has reduced maximum delay because its equivalent CP size is half the size of the other frames.}

\subsection{Complexity of FFT-based Receiver}
The complexity of the \textit{Integer Grid} estimator is dictated by the channel estimation operation shown in the Subsection \ref{subsec:2d_channel_estimation}.
\rev{And the \textit{Fine Grid} estimator has additional costs to compute \eqref{eq:dd_estimation}.}

\rev{\subsubsection{Channel Estimation, Subsection \ref{subsec:2d_channel_estimation}} Taking the example of OFDM processing, the channel estimation requires $2 |\mathcal{M}|$ \acp{FFT} of size $|\mathcal{K}|$, $|\mathcal{K}||\mathcal{M}|$ complex-valued multiplications to equalize the channel, and then $|\mathcal{K}|$ \acp{FFT} of size $|\mathcal{M}|$.
\subsubsection{Fine Grid Estimation, equation \eqref{eq:dd_estimation}} to achieve a resolution of $1/N_{\rm grid}$ about the integer grid estimates, the fine grid estimator \eqref{eq:dd_estimation} computes $|\mathcal{K}| |\mathcal{M}|$ complex-valued multiplications $N_{\rm fg} = 2 N_{\rm iterations} P  \log_2 N_{\rm grid}$ times for all targets and iterations of Algorithm 1.
$ N_{\rm iterations} P$ are the number of outer iterations and targets. 
$2 \log_2 N_{\rm grid}$ is because of a binary grid search that subdivides the grid by half in each step, and each step tests 2 candidates.}

\rev{Taking $|\mathcal{K}|$ and $|\mathcal{M}|$ from \eqref{eq:K_set} and \eqref{eq:M_set}, respectively, considering that the $K$-size FFT consumes $K \log_2 K$, the above complexities in terms fo complex-valued multiplications are  $O_{\rm ofdm} = 2 M K \log_2 K  + KM + K M \log_2 M + N_{\rm fg} KM = MK(\log_2 MK^2+1+N_{\rm fg})$, $O_{\rm ps} = M_{\rm p}K(\log_2 M_{\rm p}K^2+1+N_{\rm fg})$, $O_{\rm uw1} = N_{\rm cp}M( \log_2 M N_{\rm cp}^2 +1+N_{\rm fg})$ and $O_{\rm uw2} = MN_{\rm cp}/2M(\log_2 (N_{\rm cp}/2)^2 +1+N_{\rm fg})$.}

%
%

Not surprisingly, the complexity scales with the number of samples to be processed, i.e., the processing gain, indicating a trade-off between processing gain and complexity.
It is worth noting that 

\subsection{SNR Loss of FFT-based Receiver}\label{subsec:radar_snr}
\begin{table}[t!]
	\centering
	\footnotesize
	\caption{SNR Loss of OFDM.}
	\vspace{-3mm}	
	\begin{tabular}{c|c|c|c|c|c}\toprule
		QAM order 	   & 4 & 16 & 64 & 256 & 1024  \\ \midrule
		$L_{\rm ofdm}$ & 1 & 1.89 & 2.68 & 3.43 & 4.17\\
		$10 \log_{10} L_{\rm ofdm}$ & 0 & 2.76 & 4.29 & 5.36 & 6.2
		\\ \bottomrule
	\end{tabular}\label{tab:snr_loss_ofdm}
\end{table}
It is convenient to work with the SNR after radar processing, which has the definition below 
\setcounter{equation}{55}
\begin{equation}\label{eq:radar_snr}
\rho = G |h|^2/\sigma_{\rm w}^2,
\end{equation}
%
for one target, where $G$ is the processing gain and is given in Subsection~\ref{subsection:processing_gain} for each frame.

For the OFDM frame, the SNR loss $L_{\rm ofdm}$ happens due to the non-constant amplitude signaling in the frequency domain.
It is computed in equation \eqref{eq:L_loss_ofdm}, and the values for pertinent QAM constellation are given in Table~\ref{tab:snr_loss_ofdm}.

For the UW1 frame, the SNR loss is caused by the CP-restoration procedure, which superposes additional noise and OFDM symbol terms to the signal of interest.
Combining \eqref{eq:radar_snr} with the SNR loss given in \eqref{eq:L_loss_uw}, we find the formula
\begin{equation}\label{eq:snr_loss_uw1}
\rev{L_{\rm uw1} = 2 +\rho/G_{\rm uw1} + \beta^2 |h_{\rm los}|^2/\sigma_{\rm w}^2.}
\end{equation}
%
As such, its \ac{SINR} is given by
\begin{equation}\label{eq:radar_snr_uw}
	\rev{\begin{split}	\rho_{{\rm uw1}} & = \rho/\left (2 + \rho/G_{\rm uw1} + \beta^2 |h_{\rm los}|^2/\sigma_{\rm w}^2\right )
		\\ & \approx  \left\{\begin{matrix}
		\rho/2, & \rho \ll  G_{\rm uw1} \,\,\, {\rm and} \,\,\, \beta^2 |h_{\rm los}|^2 \ll \sigma^2_{\rm w}
		\\ 	G_{\rm uw1}, & \rho \gg G_{\rm uw1} \,\,\, {\rm and} \,\,\, \beta^2 |h_{\rm los}|^2 \ll \sigma^2_{\rm w}
	\end{matrix}\right.,
	\end{split}}
\end{equation}
where the second line is the asymptotic SINR without LoS interference in high and low SNR regimes.
%
%
In particular, we can observe that in the low SNR regime, the SNR loss for the UW1 frame is 2.
In the high SNR region, the loss scales proportionally with the SNR before the radar processing.
As a result, the maximum SINR after radar processing is $G_{\rm uw1}$. 

\section{Numerical Results}\label{sec:numerical_results}
\begin{figure*}
	\input{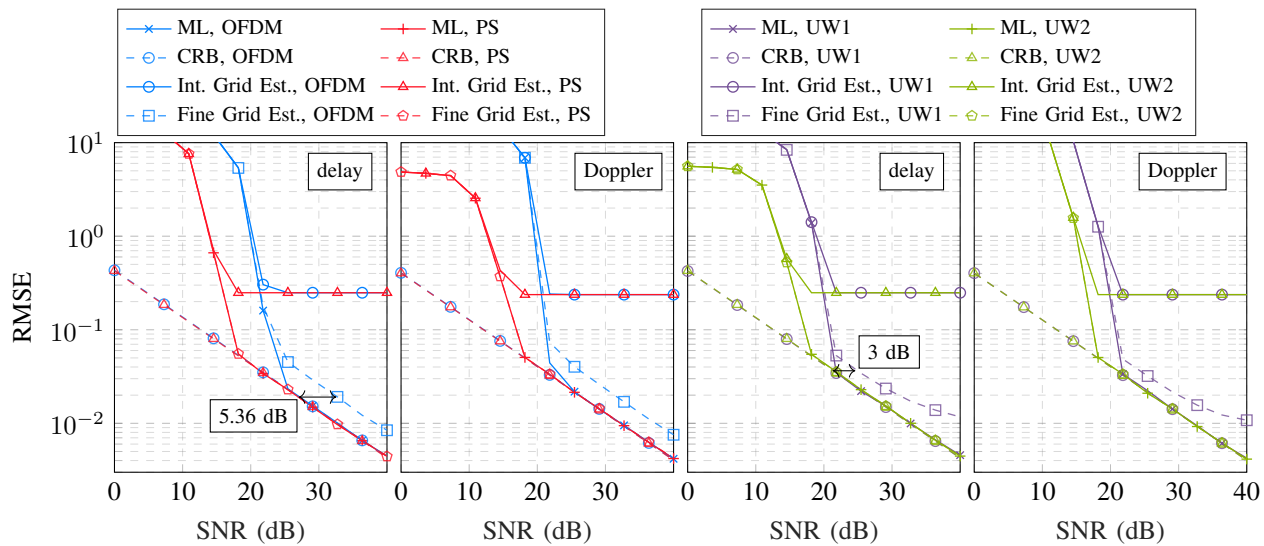} 
	\caption{Analysis of the \rev{single-target} DD estimators of Section \ref{sec:delay_doppler_estimation} \rev{without LoS interference}. \rev{The delay and Doppler are fixed, $\tilde{\tau} = 4.249$ and $\tilde{\nu} = 2.237$.}}
	\label{fig:delay_doppler_estimation}
	\vspace{-0.2cm}
\end{figure*}

\begin{table}[t!]
	\centering
	\footnotesize
	\caption{Simulation Parameters for Figs.~\ref{fig:delay_doppler_estimation}, \ref{fig:delay_los_removal}, \ref{fig:multi_target} and \ref{fig:outlier_probability}.}
	\vspace{-3mm}	
	\begin{tabular}{c|c|c|c}\toprule
		Param. & Value & Param. & Value  \\ \midrule
		FFT size, $K$ & \rev{128} &  CP size, $N_{\rm cp}$ & \rev{32} \\
		n. of sub-vectors, $M$ & 64 & 	OFDM QAM order & 256\\ 
		n. of PS sub-vectors, $M_{\rm p}$ & 8 & roll-off factor, $\alpha$ &  0.25 \\\bottomrule
	\end{tabular}\label{tab:simulation_parameters_param_estimation}
	\vspace{-0.2cm}
\end{table}

\subsection{Delay-Doppler Estimation}
In this subsection, we show the results related to the DD estimators shown in Section \ref{sec:delay_doppler_estimation}.
\rev{In particular, the \textit{Fine Grid} and \textit{Integer Grid} estimators are evaluated via Monte Carlo simulations, which are described in Subsections  \ref{subsec:fine_grid_estimation} and \ref{subsec:integer_grid}, respectively. For benchmarking, the MLE of Subsection \ref{subsec:MLE} and the CRB of Subsection \ref{subsec:CRB} are evaluated.}
The parameters for this numerical evaluation are shown in Table~\ref{tab:simulation_parameters_param_estimation}.
The \emph{Fine Grid} estimator described in Algorithm 1 divides the \emph{Integer Grid} into 256 parts to compute \eqref{eq:dd_estimation} and \eqref{eq:dd_estimation_los} which is achieved by 8 binary refinement steps, which is enough resolution to guarantee a \ac{RMSE} normalized to the DD resolution under $10^{-2}$ at high SNR.
\rev{The estimators are compared with the same radar SNR defined in \eqref{eq:radar_snr} to allow an evaluation of the SNR loss of Table~\ref{tab:frame_design_overview}.
This means that the effect of the processing gain, or number of samples processed, is not observed in these plots.	
A high-level analysis is done in Subsection \ref{subsec:high_level_analysis} that takes into consideration the radar processing gain effects in terms of range.}

\rev{Fig.~\ref{fig:delay_doppler_estimation} evaluates the RMSE of the radar normalized DD pair one target and $\beta=0$, which are fixed to $\tilde{\tau} = 4.249$ and $\tilde{\nu} = 2.237$ to allow the CRB with the unknown fixed parameter.}
We first note that the MLEs meet the CRBs in a high SNR regime, indicating their accuracy.
Moreover, the results reveal that UW2 and PS frames with the \emph{Fine Grid} estimator achieve the ML.
The reason is that the equalization does not incur loss nor does it change the noise statistics due to constant amplitude.
For OFDM, we note that the SNR loss is $10 \log_{10} L_{\rm ofdm} = 5.36$ for 256 QAM according to Table~\ref{tab:snr_loss_ofdm} when the \emph{Fine Grid} estimator is employed.
For the UW1 case, the extreme cases described by equation \eqref{eq:radar_snr_uw} are observed, namely, in the low radar SNR region, the SNR loss is 3 dB about the UW2 case.
In the high SNR region, SINR stagnates at $\rev{G_{\rm uw1} = N_{\rm cp} M = 2048}$. 
This can be observed in Fig.~\ref{fig:delay_doppler_estimation}, where UW1 frame with the \emph{Fine Grid} estimator displays a floor at high SNR.
We also note that the waterfall region of OFDM and UW1 frames is shifted to the right.
Even though their respective MLEs achieve the CRB, it happens at a higher SNR than the PS and UW2 frames, indicating their fundamental suboptimality.
Lastly, the \emph{Integer Grid} estimator achieves an RMSE exactly matching the fractional delay and Doppler which are $2.49 \times 10^{-1}$ and $2.37 \times 10^{-1}$, respectively.
This indicates that the \emph{Integer Grid} estimator estimates the integer DD with high probability.
A more detailed assessment in this regard is provided in the subsequent subsection.

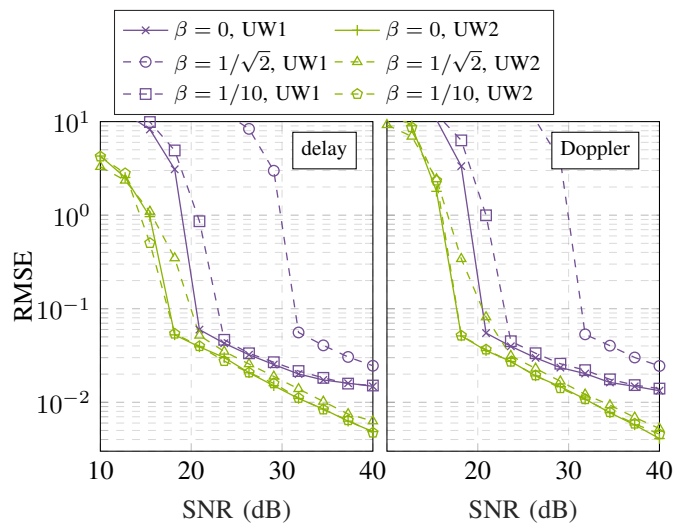
\begin{figure}[t!]
%
%
\definecolor{mycolor1}{rgb}{0.00000,0.44700,0.74100}%
\definecolor{mycolor2}{rgb}{0.85000,0.32500,0.09800}%
\definecolor{mycolor3}{rgb}{0.92900,0.69400,0.12500}%
\definecolor{mycolor4}{rgb}{0.49400,0.18400,0.55600}%
\begin{tikzpicture}
	
	\begin{semilogyaxis}[%
		scale = 0.6,
		width=3.0in,
		height=3.5in,
		at={(3.0in,0.0in)},
	xmin=10,
xmax=40,
xlabel style={font=\color{white!15!black}},
xlabel={SNR (dB)},
ymin=0.003,
ymax=10,
yminorticks=true,
ylabel={RMSE},
ylabel style={yshift=-0.2cm},
grid = both,
xticklabels={,10,20,30,40},
major grid style = {
	dashed,
	gray,
	line width = 0.25,
	opacity = 0.3,
},
major tick style = {																			
	line width = 0.25pt,
	major tick length = 3pt,
	opacity = 0.3,
},		
minor grid style = {
	dashed,
	gray,	
	line width = 0.25,
	opacity = 0.3,
},
minor tick style = {																			
	line width = 0.25pt,
	minor tick length = 3pt,
	opacity = 0.3,																																		
},
		legend style={at={(0.05,1.02)}, anchor=south west, legend cell align=left, align=left, draw=white!15!black,legend columns=2}
		]
		
		\addplot [color=darklavender, line width=0.5pt, mark=x, mark repeat={1},mark phase=1]
table[row sep=crcr]{
	10   13.796596\\12.727273   12.299289\\15.454545   8.2419817\\18.181818   3.0945871\\20.909091   0.059777208\\23.636364   0.042743229\\26.363636   0.032214883\\29.090909   0.02532389\\31.818182   0.020062368\\34.545455   0.017326474\\37.272727   0.015993025\\40   0.014814086\\
};
			\addlegendentry{\footnotesize $\beta=0$, UW1}
			
					\addplot [color=applegreen, line width=0.5pt,mark = +, mark repeat={1},mark phase=1]
table[row sep=crcr]{
	10   4.2866884\\12.727273   2.5283229\\15.454545   0.96490485\\18.181818   0.052573359\\20.909091   0.039133892\\23.636364   0.030067028\\26.363636   0.020694267\\29.090909   0.015037246\\31.818182   0.010988044\\34.545455   0.0084723948\\37.272727   0.006286662\\40   0.0048591511\\
};
			\addlegendentry{\footnotesize $\beta=0$, UW2}
		
				\addplot [color=darklavender, dashed, line width=0.5pt, mark=o, mark options={solid},opacity=1, mark repeat={1}]
table[row sep=crcr]{
	10   14.270135\\12.727273   14.469025\\15.454545   14.9297\\18.181818   14.399236\\20.909091   13.89822\\23.636364   12.525879\\26.363636   8.3648646\\29.090909   2.984725\\31.818182   0.055739123\\34.545455   0.040578506\\37.272727   0.030518584\\40   0.024570607\\
};
		\addlegendentry{\footnotesize $\beta=1/\sqrt{2}$, UW1}
		
				\addplot [color=applegreen, dashed, line width=0.5pt, mark=triangle, mark options={solid},opacity=1, mark repeat={1}]
table[row sep=crcr]{
	10   3.293686\\12.727273   2.3523191\\15.454545   1.0929242\\18.181818   0.34785422\\20.909091   0.051839679\\23.636364   0.034917382\\26.363636   0.025650977\\29.090909   0.018909938\\31.818182   0.013906518\\34.545455   0.01020155\\37.272727   0.0074986587\\40   0.0062891504\\
};
				\addlegendentry{\footnotesize $\beta=1/\sqrt{2}$, UW2}

\addplot [color=darklavender, dashed, line width=0.5pt,mark = square,mark options = {solid},opacity=1, mark repeat={1}]
table[row sep=crcr]{
	10   14.072686\\12.727273   12.990701\\15.454545   9.8985963\\18.181818   4.9191029\\20.909091   0.8612965\\23.636364   0.046514158\\26.363636   0.033395399\\29.090909   0.026817234\\31.818182   0.021574219\\34.545455   0.018247711\\37.272727   0.01584254\\40   0.015105212\\
};
\addlegendentry{\footnotesize $\beta =1/10$, UW1}

		\addplot [color=applegreen, dashed, line width=0.5pt,mark = pentagon,mark options = {solid},opacity=1, mark repeat={1}]
table[row sep=crcr]{
	10   4.2315008\\12.727273   2.7986562\\15.454545   0.50313431\\18.181818   0.054090164\\20.909091   0.039815418\\23.636364   0.027550538\\26.363636   0.02078897\\29.090909   0.015920913\\31.818182   0.011120346\\34.545455   0.0083645934\\37.272727   0.0064253657\\40   0.0047318211\\
};
		\addlegendentry{\footnotesize $\beta=1/10$, UW2}
			
	\end{semilogyaxis}

	\begin{semilogyaxis}[%
		scale = 0.6,
		width=3.0in,
		height=3.5in,
		at={(4.5in,0.0in)},
		xmin=10,
	xmax=40,
	xlabel style={font=\color{white!15!black}},
	xlabel={SNR (dB)},
	ymin=0.003,
	ymax=10,
	yminorticks=true,
	grid = both,
	yticklabels={,,},
	xticklabels={,,20,30,40},
	major grid style = {
		dashed,
		gray,
		line width = 0.25,
		opacity = 0.3,
	},
	major tick style = {																			
		line width = 0.25pt,
		major tick length = 3pt,
		opacity = 0.3,
	},		
	minor grid style = {
		dashed,
		gray,	
		line width = 0.25,
		opacity = 0.3,
	},
	minor tick style = {																			
		line width = 0.25pt,
		minor tick length = 3pt,
		opacity = 0.3,																																		
	},
		]

		\addplot [color=darklavender, line width=0.5pt, mark=x, mark repeat={1},mark phase=1]
table[row sep=crcr]{
	10   17.096874\\12.727273   16.072228\\15.454545   10.460622\\18.181818   3.3403321\\20.909091   0.054945779\\23.636364   0.03968972\\26.363636   0.029874355\\29.090909   0.023467696\\31.818182   0.020269176\\34.545455   0.016283141\\37.272727   0.014801292\\40   0.01329446\\
};

\addplot [color=applegreen, line width=0.5pt,mark = +, mark repeat={1},mark phase=1]
table[row sep=crcr]{
	10   14.151675\\12.727273   9.0204081\\15.454545   1.7979101\\18.181818   0.051245973\\20.909091   0.035701907\\23.636364   0.027318626\\26.363636   0.019149749\\29.090909   0.014693075\\31.818182   0.010804022\\34.545455   0.0077414031\\37.272727   0.0057286221\\40   0.0040976634\\
};

\addplot [color=darklavender, dashed, line width=0.5pt, mark=o, mark options={solid},opacity=1, mark repeat={1}]
table[row sep=crcr]{
	10   18.65017\\12.727273   18.142064\\15.454545   18.538933\\18.181818   18.240921\\20.909091   17.010829\\23.636364   15.469647\\26.363636   10.332153\\29.090909   3.8538475\\31.818182   0.053245185\\34.545455   0.040286239\\37.272727   0.030109868\\40   0.024451153\\
};

\addplot [color=applegreen, dashed, line width=0.5pt, mark=triangle, mark options={solid},opacity=1, mark repeat={1}]
table[row sep=crcr]{
	10   9.2271942\\12.727273   6.9401926\\15.454545   2.4313222\\18.181818   0.34142071\\20.909091   0.081404683\\23.636364   0.031437815\\26.363636   0.023007038\\29.090909   0.016777522\\31.818182   0.012148693\\34.545455   0.0093305059\\37.272727   0.0068967198\\40   0.0052402972\\
};

\addplot [color=darklavender, dashed, line width=0.5pt,mark = square,mark options = {solid},opacity=1, mark repeat={1}]
table[row sep=crcr]{
	10   17.30728\\12.727273   16.189685\\15.454545   12.033712\\18.181818   6.3007874\\20.909091   0.99675486\\23.636364   0.044935833\\26.363636   0.033441774\\29.090909   0.025824908\\31.818182   0.022051994\\34.545455   0.017591141\\37.272727   0.015261996\\40   0.014012012\\
};

\addplot [color=applegreen, dashed, line width=0.5pt,mark = pentagon,mark options = {solid},opacity=1, mark repeat={1}]
table[row sep=crcr]{
	10   13.990577\\12.727273   8.6825376\\15.454545   2.3359531\\18.181818   0.051510979\\20.909091   0.036589476\\23.636364   0.027115056\\26.363636   0.019372726\\29.090909   0.014122204\\31.818182   0.010774948\\34.545455   0.0078343886\\37.272727   0.0059041518\\40   0.0045535933\\
};
	\end{semilogyaxis}
	
	\draw[rounded corners = 0pt	] (10.6,4.0)
	node[fill = white, opacity = 1,draw=black,
	]{\footnotesize delay};
	\draw[rounded corners = 0pt	] (14.2,4.0)
	node[fill = white, opacity = 1,draw=black,
	]{\footnotesize Doppler};
	
%
%
	
\end{tikzpicture}%
\vspace{-0.3cm} 
	\vspace{-0.3cm}
	\caption{\rev{RMSE with different values of $\beta$. The target DD are distributed as $\tilde{\tau} \sim \mathcal{U}(3.5,4.5)$, $\tilde{\nu} \sim \mathcal{U}(1.5,2.5)$, $\tilde{\tau}_{\rm los} \sim \mathcal{U}(0,0.5)$ and $\tilde{\nu}_{\rm los} \sim \mathcal{U}(0,0.1)$.}}
	\label{fig:delay_los_removal}
	\vspace{-0.3cm}
\end{figure}

\rev{Fig.~\ref{fig:delay_los_removal} evaluates the immunity that the UW1 and UW2 frames have concerning the LoS interference with a single target.
	The results show the RMSE of the difference $\hat{\tilde{\tau}}_{{\Delta}_p} = \hat{\tilde{\tau}}_p - \hat{\tilde{\tau}}_{\rm los}$ and $\hat{\tilde{\nu}}_{{\Delta}_p} = \hat{\tilde{\nu}}_p - \hat{\tilde{\nu}}_{\rm los}$, so that TO and CFO are compensated.
	This evaluation considers the LoS radar SNR of $50 \, {\rm dB}$, so that the LoS DD can be accurately estimated and is typically orders of magnitude larger than the target radar SNR.
	Three antenna rejection conditions are considered, namely, 	$\beta = \left\{0,1/10,1/\sqrt{2} \right\}$. 
	The target delay and Doppler follows the uniform distribution $\tilde{\tau} \sim \mathcal{U}(3.5,4.5)$ and $\tilde{\nu} \sim \mathcal{U}(1.5,2.5)$, respectively. 
	The LoS delay and frequency shift follow $\tilde{\tau}_{\rm los} \sim \mathcal{U}(0,0.5)$ and $\tilde{\nu}_{\rm los} \sim \mathcal{U}(0,0.1)$, respectively.
	The results reveal a high sensitivity of the UW1 system regarding the LoS interference when the antenna rejection is low. 
	Specifically, for  $\beta = 1/\sqrt{2}$ we have the term\footnote{\rev{The LoS radar SNR is $50 \, {\rm dB}$, meaning that $G_{\rm uw1} |h_{\rm los}|^2/\sigma_{\rm w}^2 = 10^5$ for $G_{\rm uw1} = N_{\rm cp} M = 2048$.}} $\beta^2 |h_{\rm los}|^2/\sigma_{\rm w}^2 = 24.41$ in \eqref{eq:snr_loss_uw1} which sufficient to cause a significant performance loss.
	On the other hand, when $\beta = 1/10$ we have $\beta^2 |h_{\rm los}|^2/\sigma_{\rm w}^2 = 0.49$ with negligible impact.
	Regarding the UW2 frame, it is considerably more resilient to high LoS interference, although it has some nonnegligible performance loss when the LoS is high, due to imperfect LoS interference removal as a product of the approximated model \eqref{eq:Y_uw2}.}

\begin{figure}[t!]
	\input{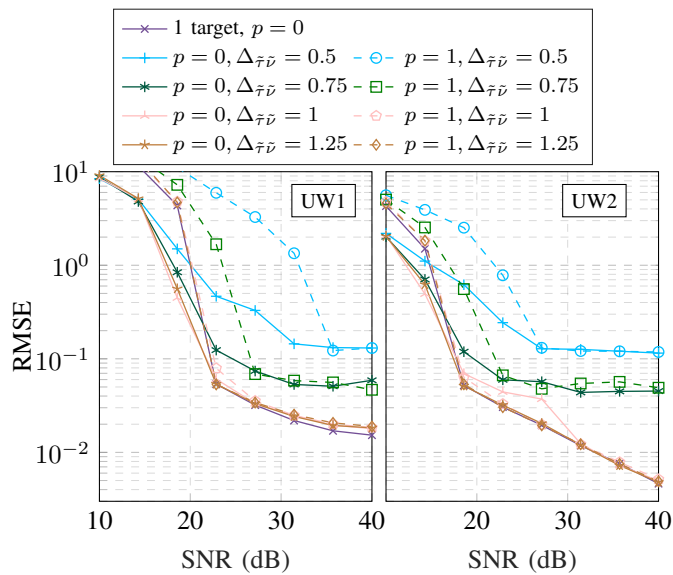} 
	\vspace{-0.3cm}
	\caption{\rev{Delay estimation for 2 targets with different fraction DD deviations $\Delta_{\tilde{\tau}\tilde{\nu}}$. The target DDs are distributed as $\tilde{\tau}_0 \sim \mathcal{U}(3.5,4.5)$, $\tilde{\nu}_0 \sim \mathcal{U}(1.5,2.5)$, $\tilde{\tau}_1 = \tilde{\tau}_0 + \Delta_{\tilde{\tau}\tilde{\nu}}$, $\tilde{\nu}_1 = \tilde{\nu}_0 + \Delta_{\tilde{\tau}\tilde{\nu}}$, $\tilde{\tau}_{\rm los} \sim \mathcal{U}(0,0.5)$ and $\tilde{\nu}_{\rm los} \sim \mathcal{U}(0,0.1)$.}}
	\label{fig:multi_target}
	\vspace{-0.3cm}
\end{figure}

\begin{figure*}
	\input{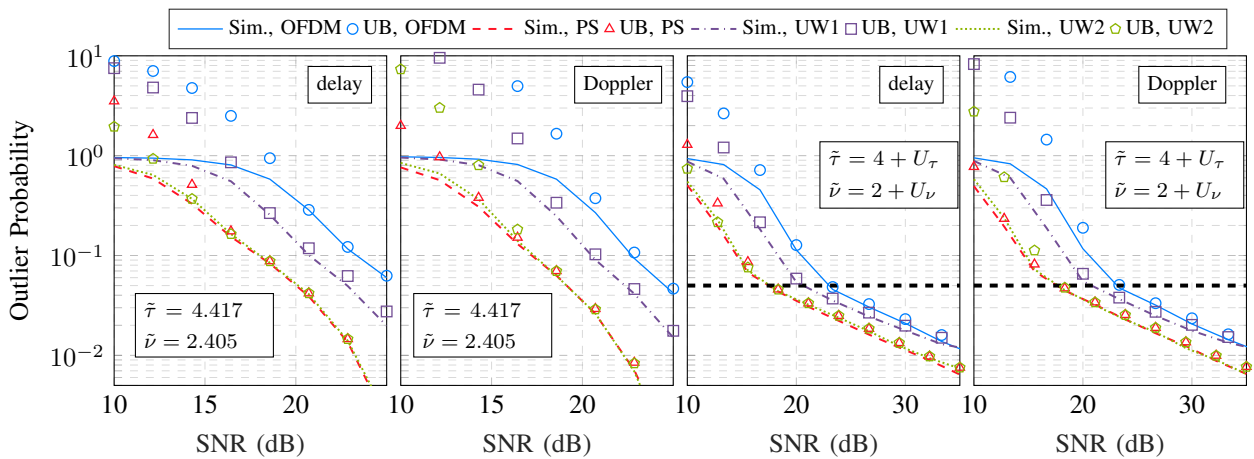} 
	\caption{\rev{Outlier probability. First and second graphs have fixed DD $(\tilde{\tau},\tilde{\nu}) = (4.417,2.405)$. Third and forth graphs have integer DD $(\tilde{\tau}_{\rm I},\tilde{\nu}_{\rm I}) = (4,2)$ and factional DD uniformly distributed between $-1/2$ and $1/2$, so that $\tilde{\tau} \sim \mathcal{U}(3.5,4.5)$, $\tilde{\nu} \sim \mathcal{U}(1.5,2.5)$.}}
	\label{fig:outlier_probability}
	\vspace{-0.2cm}
\end{figure*}

\rev{The multi-target delay difference estimation is evaluated in Fig.~\ref{fig:multi_target} with moderate LoS interference $\beta=1/10$ with the same LoS model as the results of Fig.~\ref{fig:delay_los_removal}. 
This simulation considers two targets with equal radar SNRs with $N_{\rm iterations}=8$ in Algorithm 1.
The first target, $p=0$, has the same DD distribution as the one of Fig.~\ref{fig:delay_los_removal}, i.e., $\tilde{\tau}_0 \sim \mathcal{U}(3.5,4.5)$, $\tilde{\nu}_0\sim \mathcal{U}(1.5,2.5)$.
The second target, $p=1$, has its DD as $\tilde{\tau}_1 = \tilde{\tau}_0 + \Delta_{\tilde{\tau}\tilde{\nu}}$ and \revv{$\tilde{\nu}_1 = \tilde{\nu}_0 + \Delta_{\tilde{\tau}\tilde{\nu}}$} for $\Delta_{\tilde{\tau}\tilde{\nu}} = \left\{0.5,0.75,1,1.25\right\}$ so that we can evaluate the multi-target estimation capability for different fractional DD deviations.
The results for $\Delta_{\tilde{\tau}\tilde{\nu}} = 0.5$ lead to a high floor in both frames, which is still lower than the one of the \emph{Integer Grid} estimator when compared to the results of Fig.~\ref{fig:delay_doppler_estimation}.
As expected, the error floor decreases systematically as the deviation $\Delta_{\tilde{\tau}\tilde{\nu}}$ increases.
For $\Delta_{\tilde{\tau}\tilde{\nu}} = 1.25$, both frames can estimate both targets as accurately as possible. 
Here we note that the UW1 does not achieve the single target benchmark curve because it has more CP-restoration interference with $P=2$ than with $P=1$, as shown in \eqref{eq:L_loss_uw}.}


\subsection{Outlier Probability}
This subsection investigates the outlier probability for \rev{one target and perfect LoS interference removal} using the parameters of Table~\ref{tab:simulation_parameters_param_estimation} to evaluate the outlier probability UBs \eqref{eq:P_delay} and \eqref{eq:P_doppler}.
The results are shown in Fig.~\ref{fig:outlier_probability} \rev{in the next page}, where the simulated outlier probability is also plotted.
\rev{The first and second graph depicts the result for a fixed value of DD $(\tilde{\tau},\tilde{\nu}) = (4.417,2.405)$, while the third and fourth graphs have a more general case with integer DD $(\tilde{\tau}_{\rm I},\tilde{\nu}_{\rm I}) = (4,2)$ and factional DD uniformly distributed between $-1/2$ and $1/2$, so that $\tilde{\tau} \sim \mathcal{U}(3.5,4.5)$, $\tilde{\nu} \sim \mathcal{U}(1.5,2.5)$.}
The UBs \eqref{eq:P_delay} and \eqref{eq:P_doppler} become tight at relatively high SNR, as expected.
They approach the simulation after the waterfall region when compared to the RMSE of Fig.~\ref{fig:delay_doppler_estimation}.
For the random fractional DD, we see that the outlier probability becomes less steep.
This happens because there are some values of fractional DD that have relatively high peaks for $k\neq k_0$ and $m \neq m_0$, which increases the overall outlier probability.

Moreover, a useful waterfall region analysis is done with these outcomes, that is a region from a given SNR reference point where the RMSE and outlier probability drop considerably.
Specifically, based on the results of Fig.~\ref{fig:outlier_probability}, we can define an SNR beyond which an outlier probability of $P_{\rm delay} \leq 5\times 10^{-2}$ and $P_{\rm Doppler} \leq 5\times 10^{-2}$ is expected, which have tight outlier probability.
For the PS and UW2 frames, this happens beyond 17 dB.
For the OFDM frame, it happens after 17 + $10 \log_{10}L_{\rm ofdm} = 22.36$ dB.
For the UW1 frame, the limit is $17 + 3.01 = 20.01$ dB, since in the relatively low SNR region, the SNR loss for the UW1 frame is approximately 3.01 dB.
These reference numbers are used in the range analysis performed in the subsequent subsection.


\subsection{Range Comparison of Different Frame Schemes}\label{subsec:high_level_analysis}

\begin{table}[t!]
	\centering
	\footnotesize
	\caption{Simulation Parameters for Fig.~\ref{fig:high_level_comparison}.}
	\vspace{-3mm}	
	\begin{tabular}{c|cc}\toprule
		Parameter & Outdoor	& Indoor \\ \midrule
		OFDM SCS & $\SI{120}{\kilo\Hz}$  &  $\SI{480}{\kilo\Hz}$ \\
		bandwidth, $B$ & $\SI{122.88}{\mega\Hz}$ &  $\SI{491.52}{\mega\Hz}$\\
		BS-radar distance, $d_{\rm BR}$   &  $\SI{200}{\meter}$ &  $\SI{20}{\meter}$\\
		radar RCS, $\sigma_{\rm rcs}$ & $\SI{10}{\metre\squared}$ & $\SI{1}{\metre\squared}$ \\ 
		FFT size, $K$ 			& \multicolumn{2}{c}{1024} \\
		number of sub-vectors, $M$ 	& \multicolumn{2}{c}{140}\\ 
		number of PS sub-vectors, $M_{\rm p}$ & \multicolumn{2}{c}{20} \\
		CP size, $N_{\rm cp}$ 	& \multicolumn{2}{c}{204} \\
		OFDM QAM order 			& \multicolumn{2}{c}{256} \\ 
		roll-off factor, $\alpha$ &  \multicolumn{2}{c}{0.25}  \\ 
		center frequency, $f_c$  	&  \multicolumn{2}{c}{$\SI{28}{\giga\Hz}$}  \\ 
		transmit power, $P_{\rm bs}$  	&  \multicolumn{2}{c}{$23.98\,\text{dBm}$}  \\ 
		path loss exponent, $\eta$ & \multicolumn{2}{c}{2.3 \cite{RappaportmmWave}} \\
		BS antenna gain, $G_{\rm bs}$ & \multicolumn{2}{c}{8} \\
		Radar antenna gain, $G_{\rm r}$ & \multicolumn{2}{c}{64} \\
		noise PSD &   \multicolumn{2}{c}{$-174 \, \text{dBm/Hz}$}\\ 
		\bottomrule
	\end{tabular}\label{tab:high_level}
	
	\vspace{-0.2cm}
\end{table}

\begin{figure}[t!]
	\input{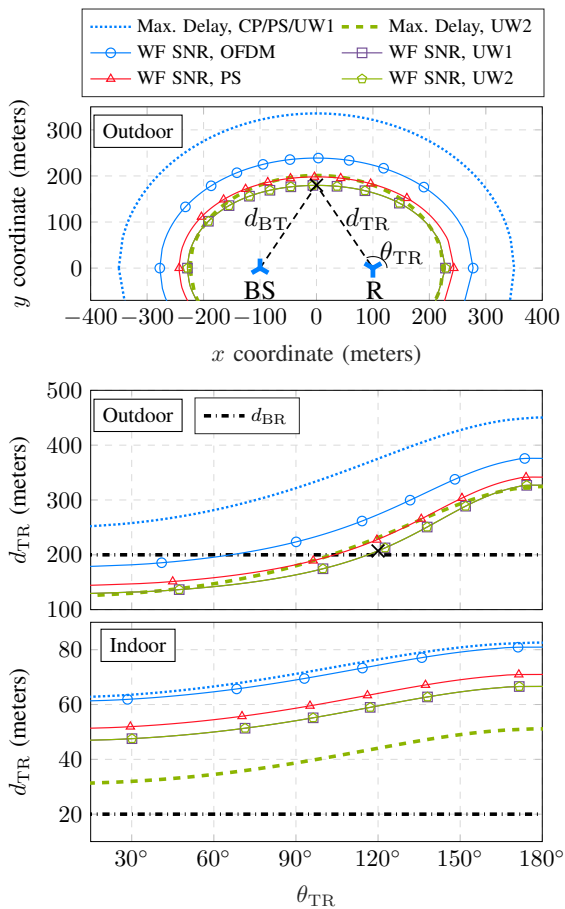} 
	\caption{Range comparison of the OFDM, PS, UW1, and UW2 frames. The top graph shows the Cassini oval and iso-range contours taking the radar as a reference. The middle and bottom graphs plot $d_{\rm TR}$ for different angles $\theta_{\rm TR}$. The waterfall (WF) SNRs for OFDM, PS, UW1, and UW2 are 22.36, 17, 20.03, and 17 dB, respectively. }
	\label{fig:high_level_comparison}
	 \vspace{-0.5cm}
\end{figure}

\rev{This subsection provides a high-level range comparison between the OFDM, PS, UW1, and UW2 frames, for one target estimation and perfect LoS interference removal.}
\rev{The target channel model considered is the double-path loss}
\begin{equation}\label{eq:h}
	\rev{	|h|^2 = \frac{P G_{\rm bs} G_{\rm r}\lambda^2\sigma_{\rm rcs}}{(4\pi)^3 (d_{\rm BT}d_{\rm TR})^{\eta}} ,}
\end{equation}
\rev{with system level parameters are shown in Table~\ref{tab:high_level}, where $d_{\rm BT}$ is the distance between the BS and target, and $d_{\rm TR}$ is the distance between the target and radar.
	An illustration is depicted in the top graph of Fig.~\ref{fig:high_level_comparison} with the bases station (BS), passive radar (PR), and target contours which are used in the range study in the remainder of this subsection.}

Two configurations are considered, namely, outdoor and indoor.
The outdoor configuration has the distance between the BS and radar $d_{\rm BR}=\SI{200}{\metre}$ and $\sigma_{\rm rcs} = \SI{10}{\metre\squared}$.
The indoor case has the distance between the BS and radar $d_{\rm BR}=\SI{20}{\metre}$ and $\sigma_{\rm rcs} = \SI{1}{\metre\squared}$.
We note that the CP and UW sizes are $20\%$ of the FFT size, which is a typical proportion. 
The CP duration $N_{\rm cp}/B$ equals \SI{1627}{\nano\second} and \SI{406}{\nano\second} for the outdoor and indoor environments, respectively.
These are practical guard interval quantities for communications with a reasonable margin since the maximum delay in outdoor and indoor environments has been reported to be around  \SI{700}{\nano\second} \cite{RappaportItWillWork} and \SI{200}{\nano\second} \cite{RappaportmmWave} at \SI{28}{\giga\Hz}, respectively.
Lastly, we note that the UW frames offer $M_{\rm p}/(M-M_{\rm p}) \times 100 = 16.67 \%$ more data than the PS frame.

In the following, we analyze the range from two relevant perspectives.
The first is the maximum bistatic range based on the iso-range contour, where the maximum unambiguous delay in Table~\ref{tab:high_level} is considered.
{The second is the constant SNR Cassini oval where the waterfall SNRs for OFDM, PS, UW1, and UW2 are 22.36, 17, 20.03, and 17 dB, respectively.}
\rev{The details related to the iso-range contour and Cassini oval are given in Appendix \ref{app:geometry}, and an illustration is depicted in the top graph of Fig.~\ref{fig:high_level_comparison} for the outdoor setting.}

\rev{The results are better compared with the middle and bottom graphs, where the distance between the target and radar, $d_{\rm TR}$, for different angles $\theta_{\rm TR}$, are plotted matching the contours depicted in the graph at the top.}
\rev{We note that the distance between the BS and radar, $d_{\rm BR}$, is plotted as black dashed line for reference.}
Firstly, we analyze the maximum unambiguous range of OFDM, PS, and UW1 in comparison to UW2.
As expected, the range reduction of UW2 is evident due to its smaller maximum delay.
In particular, for the outdoor configuration and $\theta_{\rm TR} = 120^{\circ}$, $d_{{\rm TR}_{\rm max}} \approx \SI{230}{\metre}$ for the UW2 frame, while it is $d_{{\rm TR}_{\rm max}} \approx \SI{375}{\metre}$ for the other frames. 
It is interesting to note that although the maximum unambiguous delay of the UW2 frame is half the others, \rev{its maximum range in typically more than half of the other methods, demonstrating that the range reduction of UW2 is not as bad as one would expect.}

Regarding the maximum distance that achieves the waterfall SNRs obtained via the Cassini oval, we observe that the difference between UW1 and UW2 frames is not very high from PS for both configurations.
For example, for $\theta_{\rm TR} = 120^{\circ}$, $d_{{\rm TR}_{\rm Cass.}} \approx \SI{207}{\metre}$ for the UW1 and UW2 frames, while it is approximately \SI{228}{\metre} and \SI{273}{\metre} for PS and OFDM, respectively.
We note that the Cassini ovals are directly dependent on the processing gain and SNR loss of Table~\ref{tab:frame_design_overview}.
Since we can expect the UW1, UW2, and PS frames to have a similar processing gain for practical configurations, the Cassini ovals between the UW1, UW2, and PS frames will be similar in general.
It is also possible that the UW1 and UW2 frames have a broader Cassini oval than the PS for smaller values of $M_{\rm p}$.
This makes both UW frames to be a promising alternative to avoid sacrificing the data rate to realize sensing tasks.
When comparing both UW frames, we see that they have very similar Cassini oval.
This happens because the $2 \times$ higher processing gain of UW1 is compensated by its $2 \times$ SNR loss.
Nevertheless, the UW1 frame suffers from an SNR floor which does not appear for the UW2 frame.
Thus, when the unambiguous range is not critical, the UW2 frame is preferred.

For the indoor scenario, the Cassini oval is way beyond the maximum bistatic range, indicating that this case is limited by the CP size and not the power.
Interestingly, the maximum bistatic range surpasses \SI{30}{\metre} for the UW2 frame.
Considering that the distance between the transmitter and the radar $d_{\rm BR} = \SI{20}{\metre}$, we can expect the maximum bistatic range to be sufficient even for the UW2 frame.
Since the UW2 frame has a steeper estimation curve according to the results of Fig.~\ref{fig:delay_doppler_estimation}, in this situation UW2 frame is more suitable than the UW1.

In summary, the UW1 and UW2 frames are competitive when compared to the PS design since they achieve similar performance without compromising the data rate for communications users.
For outdoor scenarios where the maximum bistatic range is more relevant, the UW1 is a more suitable solution than the UW2.
In indoor scenarios where the range is not critical, the UW2 frame is a more suitable solution when the \emph{Fine Grid} FFT-based receiver is employed, since it does not have the error floor of the UW1 frame.

\section{Conclusion}\label{sec:conclusion}
In this work, we have proposed two solutions for ISAC frame design for bistatic radar within mobile communication systems.
Specifically, the idea consists of placing deterministic signals, based on the ZC sequence, into the guard interval instead of the CP.
The first is the UW1 with the CP size.
The second is two UWs with half of the CP size appended after each other, which is termed UW2.
The performance of the proposed UW-based frames has been compared to OFDM and PS frames.
Against OFDM, the UW frames are advantageous because they do not require the radar to know the data symbols, which is a requirement that can be impractical.
Concerning the PS, the UW solutions do not compromise the data rate for communication.

We have derived relevant theoretical results which have been validated by numerical analysis.
Considering a band-limited system with an RRC filter, we have derived the CRB.
In addition, we have derived an FFT-based \rev{multi-target} delay-Doppler (DD) estimator with \rev{LoS interference removal}, and the outlier probability computation when the integer part of the DD is erroneously estimated.
Lastly, we have compare all methods in terms of sensing range.
The UW2 has the drawback of decreasing the maximum unambiguous estimation range from 50\% to 70\%, which can be critical in outdoor sensing, but not so much for the indoor setting considered.
In terms of the maximum range due to radar SNR, both UW waveforms have a very similar range to the PS system.
In summary, the UW frames are good alternative candidates for bistatic ISAC, where the UW1 is more suitable for outdoor sensing, and the UW2 is more suitable for indoor.
  
There are several aspects to be further explored in future research, namely, i) explore the impact of clutter on the UW frames, ii) extend the model to perform angle-of-arrival estimation, which is also a typical parameter to be sensed, iii) explore the UW for line-of-sight synchronization, iv) enhance the unambiguous range of the UW2 method, \revv{and v) investigate the impact of multipath per target in the radar channel model}.

\appendix

\subsection{\rev{Discrete Time Signal Approximation}}\label{ap_subsec:discrete_time_signal}
\rev{Considering only one target and no LoS interference for simplicity of the exposition, i.e., $\beta = 0$ and $P=1$ in \eqref{eq:r}, the sampled OFDM signal is given by} 
\begin{align}\label{eq:Y_ofdm_2}
		& \rev{\mathbf{Y}_{\rm ofdm}[k,m]  = y_{\rm t}(I_{\rm ofdm}(k,m)T_{\rm s})}
		\nonumber\\ & \rev{\stackrel{(a)}{\approx} h e^{-j 2 \pi \nu I(k,m)T_{\rm s}}	\sum_{n=0}^{N_{\rm x}-1}\!\!\mathbf{x}[n]g_{\rm RC}((I(k,m)-n)T_{\rm s}-\tau)	}
		\nonumber\\ & \,\,\,\, \rev{+ w_{\rm t}'(I(k,m)T_{\rm s})}
		\nonumber
		\\ & \rev{\stackrel{(b)}{\approx} h e^{-j 2 \pi \nu I(k,m) T_{\rm s}} (\mathbf{x}_m \circledast \mathbf{g}_{K,\tilde{\tau}})[k] + \mathbf{W}[k,m]}
		\nonumber\\ & \rev{\stackrel{(c)}{\approx} h' \mathbf{u}_{M,\tilde{\nu}}[m] (\mathbf{x}_m \circledast \mathbf{g}_{K,\tilde{\tau}})[k] + \mathbf{W}[k,m],}
	\end{align}	
	\rev{for $0\leq k < K$ and $0\leq m < M$ such that $\mathbf{Y}_{\rm ofdm} \in \mathbb{C}^{K\times M}$.
	The index mapper $I_{\rm ofdm}(k,m) = (k+N_{\rm cp} + m(N_{\rm cp} + K))$ is used to place the received signal associated with the $m$th sub-block in the $m$th column of $\mathbf{Y}_{\rm ofdm}$, where the CP is removed to allow frequency domain processing.}
	\rev{In \eqref{eq:Y_ofdm_2}, $(a)$ represents the same approximation of \eqref{eq:y}.
	Line $(b)$ writes a circular convolution between the discrete time signal $\mathbf{x}$ and the raise-cosine filter normalized to the sampling interval}
	\begin{equation}\label{eq:g_K}
		\rev{\mathbf{g}_{K,\tilde{\tau}}[k] = \tilde{g}_{\rm RC}(k-\tilde{\tau}),}
	\end{equation}	
	\rev{for $k=0,1,\cdots,K-1$, where $\tilde{g}_{\rm RC}(t) = {g}_{\rm RC}(t/T_{\rm s})$, $\tilde{\tau} = \tau/T_{\rm s}$ is the delay normalized to the sampling interval, which is assumed to be in the interval $0\leq \tilde{\tau} < N_{\rm cp}$.
	The subscript $K$ in \eqref{eq:g_K} denotes its size.
	It is an approximation because the time-domain filter does not provoke a perfect circular convolution due to non-zero samples being larger than the CP size.
	Lastly, the approximation in $(c)$ considers that the term $e^{-j 2 \pi \nu I(k,m) T_{\rm s}}$ remains approximately constant within the period of the $m$th sub-block with phase}
	\begin{equation}\label{eq:u_M}
		\rev{\mathbf{u}_{M,\tilde{\nu}}[m] = \exp(-j 2 \pi \tilde{\nu}m/M),}
	\end{equation}
	\rev{where $\tilde{\nu} = \nu T_{\rm s} N_{\rm x}$ is the frequency shift normalized to subcarrier spacing of the whole signal if the bandwidth is divided by $N_{\rm x}$.
	In practice, $(c)$ tends to be less accurate as $K$ increases because it could lead to significant phase changes within a sub-block.
	The channel coefficient denoted by $h'$ takes into account the appropriate initial phase. }
	
		\rev{The approximation shown in \eqref{eq:Y_ofdm} is obtained by combining \eqref{eq:Y_ofdm_2} with the multiple targets and LoS interference of \eqref{eq:r}.}

\subsection{2D Channel Estimation}\label{ap_subsec:2D_channel_estimation}
In the following, we show that $\hat{\mathbf{H}}_{\rm ofdm}\in \mathbb{C}^{M\times K}$ of \eqref{eq:H_ofdm} holds.
\rev{Neglecting the LoS interference, and start with the single target assumption, we have}
\begin{equation}\label{eq:H_appe}
	\begin{split}
		\hat{\mathbf{H}}_{\rm ofdm}  = \mathbf{F}_M^{\rm H}(\mathbf{F}_K^{\rm H}((\mathbf{F}_K\mathbf{Y}_{\rm ofdm})\oslash \mathbf{D}))^T
		\approx h'  \mathbf{v}_{M,\tilde{\nu}} \mathbf{g}_{K,\tilde{\tau}}^{\rm T}\! +\! \mathbf{W}'.
	\end{split}
\end{equation}
Let $\pmb{g}_{K,M,\tilde{\tau}} = [\mathbf{g}_{K,\tilde{\tau}} \,\,\, \mathbf{g}_{K,\tilde{\tau}} \cdots \mathbf{g}_{K,\tilde{\tau}}] \in \mathbb{R}^{K \times M}$ be the normalized raised cosine filter replicated $M$ times along columns, and  $\pmb{u}_{M,K,\tilde{\nu}} = [\mathbf{u}_{M,\tilde{\nu}} \,\,\, \mathbf{u}_{M,\tilde{\nu}} \cdots \mathbf{u}_{M,\tilde{\nu}}] \in \mathbb{R}^{M \times K}$ be the Doppler phase shift replicated $K$ times along columns.
Ignoring the noise component and the channel phase $h'$ in \eqref{eq:Y_ofdm}, we first note that $\mathbf{F}_K\mathbf{Y}_{\rm ofdm} \approx (\pmb{u}_{M,K,\tilde{\nu}})^{\rm T} \odot \pmb{G}_{K,M,\tilde{\tau}} \odot \mathbf{D} $, where $\pmb{G}_{K,M,\tilde{\tau}} = \mathbf{F}_K\pmb{g}_{K,M,\tilde{\tau}}$ and $\mathbf{D}$ are the data symbols stacked column-wise for each transmitted sub-vector.
Notice that this can be done because the rows of $(\pmb{u}_{M,K,\tilde{\nu}})^{\rm T}$ are the same for a particular column.
Then, the element-wise division by the data leads to $(\mathbf{F}_K\mathbf{Y}_{\rm ofdm})\oslash \mathbf{D} \approx  (\pmb{u}_{M,K,\tilde{\nu}})^{\rm T} \odot \pmb{G}_{K,M,\tilde{\tau}}$, removing the data symbol.
After that, the \ac{IDFT} $\mathbf{F}_K^{\rm H}((\mathbf{F}_K\mathbf{Y})\oslash \mathbf{D}) \approx  (\pmb{u}_{M,K,\tilde{\nu}})^{\rm T} \odot \pmb{g}_{K,M,\tilde{\tau}}$ recovers back the raise cosine filter in time domain. 
Since the Doppler shift is encoded in the linear phase shift of $\mathbf{u}_{M,\tilde{\nu}}$, the phase shift can be estimated by the IDFT of size $M$, which is achieved by transposing the previous result and taking the IDFT as $\mathbf{F}_M^{\rm H}(\mathbf{F}_K^{\rm H}((\mathbf{F}_K\mathbf{Y}_{\rm ofdm})\oslash \mathbf{D}))^T \approx \mathbf{F}_M^{\rm H} (\pmb{u}_{M,K,\tilde{\nu}}) \odot (\pmb{g}_{K,M,\tilde{\tau}})^{\rm T} =  (\pmb{v}_{M,K,\tilde{\nu}}) \odot (\pmb{g}_{K,M,\tilde{\tau}})^{\rm T} = \mathbf{v}_{M,\tilde{\nu}} \mathbf{g}_{K,\tilde{\tau}}^{\rm T}$ has size $M\times K$, where $\pmb{v}_{M,K,\tilde{\nu}} = \mathbf{F}_M^{\rm H}\pmb{u}_{M,K,\tilde{\nu}}$ and $\mathbf{v}_{M,\tilde{\nu}} = \mathbf{F}_M^{\rm H}\mathbf{u}_{M,\tilde{\nu}}$.

\rev{The multiple target extension follows directly from \eqref{eq:H_appe} since the model is additive.}

\subsection{Bistatic Radar Geometry}\label{app:geometry}
Let $P_{\rm bs} = (x_{\rm bs},y_{\rm bs})$, $P_{\rm t} = (x_{\rm t},y_{\rm t})$ and $P_{\rm r} = (x_{\rm r},y_{\rm r})$ be the coordinates of the BS, target and radar, respectively, in the $x \times y$ plane.
The distances BS-target, target-radar and BS-radar are given doted by $d_{\rm BT}$, $d_{\rm TR}$ and $d_{\rm BR}$, respectively.
Without loss of generality, we place the BS and radar at $P_{\rm bs} = (-d_{\rm BR}/2,0)$ and $P_{\rm r} = (d_{\rm BR}/2,0)$.

\subsubsection{Maximum Delay Iso-range Contour}
\rev{Taking the LoS path as a reference}, the target delay is given by $\tau = (d_{\rm BT}+d_{\rm TR}-d_{\rm BR})/c$, we $c$ is the speed of light.
The maximum unambiguous delay $\tau^{\rm max}$ of the Subsection \ref{subsec:maximum_unambiguous_range} determines the maximum distance $d_{\rm max}  = \tau^{\rm max}c+d_{\rm BR} \geq d_{\rm BT}+d_{\rm TR}$ that the signal from the BS to the radar can travel, such that the $\tau \leq \tau^{\rm max}$.
The iso-range contour is defined by an ellipse that describes the target's coordinates that satisfy $d_{\rm max} = d_{\rm BT}+d_{\rm TR}$, whose coordinates are 
\begin{align}
 x_{\rm t} &=\frac{d_{\rm max}}{2} \cos(t) \nonumber\\  y_{\rm t} &= 1/2\sqrt{(\tau^{\rm max}c)^2 + 2 d_{\rm BR}\tau^{\rm max}c } \sin(t) & \text{for $0 \leq t < 2\pi$}
\end{align}
for $0 \leq t < 2\pi$.

\subsubsection{Waterfall SNR Cassini Oval}\label{app:cassini_oval}
%
For a constant $b>0$, the Cassini oval defines a set of coordinates of $\mathcal{P}_{\rm t}$ which satisfies $d_{\rm BT} d_{\rm TR} = b^2$.
In the bistatic radar scenario, the Cassini oval is useful to determine the coordinates $P_{\rm t}=(x_{\rm t},y_{\rm t})$ which attain a given desired SNR $\rho^\star$ for the channel model of \eqref{eq:h}.
In this case, the constant $b^2$ is
\begin{equation}\label{eq:b2}
b^2 = d_{\rm BT} d_{\rm BT} = \left(\frac{P_{\rm bs} G G_{\rm bs} G_{\rm r} \lambda^2 \sigma_{\rm rcs}}{\rho^\star \tilde{\sigma}_{\rm w}^2 (4\pi)^3}\right)^{\frac{1}{\eta}}.
\end{equation}
%
From \cite{cassini}, the coordinates $P_{\rm t}$ satisfying \eqref{eq:b2} are
\begin{equation}\label{eq:xy_cassini}
x_{\rm t} = \frac{b^4-t^4}{4d_{\rm BR}t^2}\,\,\,  \text{   and    } \,\,\, y_{\rm t} = \pm\sqrt{t^2- (x-d_{\rm BR})^2},
\end{equation}
for $d_{\rm BR} + \sqrt{d_{\rm BR}^2-b^2} \leq t \leq d_{\rm BR} + \sqrt{d_{\rm BR}^2+b^2}$ if $d_{\rm BR}\geq b$, and $-d_{\rm BR} + \sqrt{d_{\rm BR}^2+b^2} \leq t \leq d_{\rm BR} + \sqrt{d_{\rm BR}^2+b^2}$ otherwise.

\bibliography{references_ha}{}

\begin{thebibliography}{10}

\bibitem{chafii2023twelve}
M.~Chafii, L.~Bariah, S.~Muhaidat, and M.~Debbah, ``Twelve scientific
  challenges for {6G}: Rethinking the foundations of communications theory,''
  {\em {IEEE Commun. Surv. Tutor.}}, vol.~25, no.~2, pp.~868--904, 2023.

\bibitem{LiuJSAC}
F.~Liu {\em et~al.}, ``Integrated sensing and communications: Toward
  dual-functional wireless networks for 6g and beyond,'' {\em {IEEE J. Sel.
  Areas Commun.}}, vol.~40, no.~6, pp.~1728--1767, 2022.

\bibitem{ZhangISAC}
J.~A. Zhang, M.~L. Rahman, K.~Wu, X.~Huang, Y.~J. Guo, S.~Chen, and J.~Yuan,
  ``Enabling joint communication and radar sensing in mobile networks—a
  survey,'' {\em {IEEE Commun. Surv. Tutor.}}, vol.~24, no.~1, pp.~306--345,
  2022.

\bibitem{BomfinSPAWC}
T.~M. Pham, R.~Bomfin, A.~Nimr, A.~N. Barreto, P.~Sen, and G.~Fettweis, ``Joint
  communications and sensing experiments using mmwave platforms,'' in {\em 2021
  IEEE 22nd International Workshop on Signal Processing Advances in Wireless
  Communications (SPAWC)}, pp.~501--505, 2021.

\bibitem{Shlezinger}
D.~Ma, N.~Shlezinger, T.~Huang, Y.~Liu, and Y.~C. Eldar, ``Joint
  radar-communication strategies for autonomous vehicles: Combining two key
  automotive technologies,'' {\em {IEEE Signal Processing Mag.}}, vol.~37,
  no.~4, pp.~85--97, 2020.

\bibitem{Zhang_ISAC}
J.~A. Zhang {\em et~al.}, ``An overview of signal processing techniques for
  joint communication and radar sensing,'' {\em IEEE Journal of Selected Topics
  in Signal Processing}, vol.~15, no.~6, pp.~1295--1315, 2021.

\bibitem{Sturm}
C.~Sturm and W.~Wiesbeck, ``Waveform design and signal processing aspects for
  fusion of wireless communications and radar sensing,'' {\em {Proc. IEEE}},
  vol.~99, no.~7, pp.~1236--1259, 2011.

\bibitem{Barneto}
C.~Baquero~Barneto {\em et~al.}, ``Full-duplex {OFDM} radar with {LTE} and {5G
  NR} waveforms: Challenges, solutions, and measurements,'' {\em {IEEE Trans.
  Microw. Theory. Tech.}}, vol.~67, no.~10, pp.~4042--4054, 2019.

\bibitem{Johnston}
J.~Johnston, L.~Venturino, E.~Grossi, M.~Lops, and X.~Wang, ``{MIMO OFDM}
  dual-function radar-communication under error rate and beampattern
  constraints,'' {\em {IEEE J. Sel. Areas Commun.}}, vol.~40, no.~6,
  pp.~1951--1964, 2022.

\bibitem{Sayed2}
S.~H. Dokhanchi, A.~N. Barreto, and G.~P. Fettweis, ``Performance analysis of
  zero-padded sequences for joint communications and sensing,'' {\em {IEEE
  Trans. Signal Process.}}, vol.~71, pp.~1725--1741, 2023.

\bibitem{LiuTSP}
F.~Liu, L.~Zhou, C.~Masouros, A.~Li, W.~Luo, and A.~Petropulu, ``Toward
  dual-functional radar-communication systems: Optimal waveform design,'' {\em
  {IEEE Trans. Signal Process.}}, vol.~66, no.~16, pp.~4264--4279, 2018.

\bibitem{Bazzi}
A.~Bazzi and M.~Chafii, ``On outage-based beamforming design for
  dual-functional radar-communication {6G} systems,'' {\em {IEEE Trans.
  Wireless Commun.}}, vol.~22, no.~8, pp.~5598--5612, 2023.

\bibitem{BAZZI_papr}
A.~Bazzi and M.~Chafii, ``On integrated sensing and communication waveforms
  with tunable {PAPR},'' {\em {IEEE Trans. Wireless Commun.}}, vol.~22, no.~11,
  pp.~7345--7360, 2023.

\bibitem{Alhil}
A.~Chowdary, A.~Bazzi, and M.~Chafii, ``On hybrid radar fusion for integrated
  sensing and communication,'' {\em {IEEE Trans. Wireless Commun.}}, pp.~1--1,
  2024.

\bibitem{Gaudio}
L.~Gaudio, M.~Kobayashi, G.~Caire, and G.~Colavolpe, ``On the effectiveness of
  otfs for joint radar parameter estimation and communication,'' {\em {IEEE
  Trans. Wireless Commun.}}, vol.~19, no.~9, pp.~5951--5965, 2020.

\bibitem{Keskin}
M.~F. Keskin, H.~Wymeersch, and A.~Alvarado, ``Radar sensing with {OTFS}:
  Embracing {ISI} and {ICI} to surpass the ambiguity barrier,'' in {\em 2021
  IEEE International Conference on Communications Workshops (ICC Workshops)},
  pp.~1--6, 2021.

\bibitem{Gong}
Z.~Gong, F.~Jiang, C.~Li, and X.~Shen, ``Simultaneous localization and
  communications with massive mimo-otfs,'' {\em {IEEE J. Sel. Areas Commun.}},
  vol.~41, no.~12, pp.~3908--3924, 2023.

\bibitem{Nuria80211}
P.~Kumari, J.~Choi, N.~González-Prelcic, and R.~W. Heath, ``{IEEE}
  802.11ad-based radar: An approach to joint vehicular communication-radar
  system,'' {\em {IEEE Trans. Veh. Technol.}}, vol.~67, no.~4, pp.~3012--3027,
  2018.

\bibitem{SundeepJump}
J.~Pegoraro {\em et~al.}, ``{JUMP}: Joint communication and sensing with
  unsynchronized transceivers made practical,'' {\em {IEEE Trans. Wireless
  Commun.}}, pp.~1--1, 2024.

\bibitem{WeiPRS}
Z.~Wei and al, ``{5G PRS}-based sensing: A sensing reference signal approach
  for joint sensing and communication system,'' {\em {IEEE Trans. Veh.
  Technol.}}, vol.~72, no.~3, pp.~3250--3263, 2023.

\bibitem{Deneire}
L.~Deneire, B.~Gyselinckx, and M.~Engels, ``Training sequence versus cyclic
  prefix-a new look on single carrier communication,'' {\em {IEEE Commun.
  Lett.}}, vol.~5, no.~7, pp.~292--294, 2001.

\bibitem{Coon}
J.~Coon, M.~Sandell, M.~Beach, and J.~McGeehan, ``{Channel and noise variance
  estimation and tracking algorithms for unique-word based single-carrier
  systems},'' {\em {IEEE Trans. Wireless Commun.}}, vol.~5, no.~6,
  pp.~1488--1496, 2006.

\bibitem{Huemer}
M.~Huemer, H.~Witschnig, and J.~Hausner, ``{Unique word based phase tracking
  algorithms for SC/FDE-systems},'' in {\em GLOBECOM '03. IEEE Global
  Telecommunications Conference}, vol.~1, pp.~70--74 Vol.1, 2003.

\bibitem{ShahabTWC}
S.~Ehsanfar, M.~Chafii, and G.~P. Fettweis, ``On {UW}-based transmission for
  {MIMO} multi-carriers with spatial multiplexing,'' {\em {IEEE Trans. Wireless
  Commun.}}, vol.~19, no.~9, pp.~5875--5890, 2020.

\bibitem{BomfinTWC}
R.~Bomfin, M.~Chafii, A.~Nimr, and G.~Fettweis, ``A robust baseband transceiver
  design for doubly-dispersive channels,'' {\em {IEEE Trans. Wireless
  Commun.}}, vol.~20, no.~8, pp.~4781--4796, 2021.

\bibitem{BomfinCPfree}
R.~{Bomfin}, M.~{Chafii}, and G.~{Fettweis}, ``A novel iterative receiver
  design for {CP}-free transmission under frequency-selective channels,'' {\em
  {IEEE Commun. Lett.}}, vol.~24, no.~3, pp.~525--529, 2020.

\bibitem{Bomfin_Globecom}
R.~Bomfin and M.~Chafii, ``Unique word-based frame design for bistatic {ISAC}
  with time-domain filtering,'' in {\em 2024 IEEE Global Communications
  Conference (GLOBECOM)}, p.~5.93, Dec. 2024.

\bibitem{AndrewBistatic}
K.~Wu {\em et~al.}, ``Sensing in bi-static {ISAC} systems with clock
  asynchronism: A signal processing perspective,'' {\em arXiv preprint}, 2024.
\newblock arXiv:2402.09048.

\bibitem{Kavous}
N.~B. Khoolenjani and K.~Khorshidian, ``On the ratio of rice random
  variables,'' {\em Journal of the Iranian Statistical Society}, vol.~8,
  pp.~61--71, 2009.

\bibitem{frances}
E.~Díaz-Francés and F.~J. Rubio, ``On the existence of a normal approximation
  to the distribution of the ratio of two independent normal random
  variables,'' {\em Statistical Papers}, vol.~54, pp.~309--323, 2013.

\bibitem{casella}
G.~Casella and R.~L. Berger, {\em Statistical Inference}.
\newblock USA: Duxbury, 2002.

\bibitem{RappaportmmWave}
Y.~Xing and T.~S. Rappaport, ``Millimeter wave and terahertz urban microcell
  propagation measurements and models,'' {\em {IEEE Commun. Lett.}}, vol.~25,
  no.~12, pp.~3755--3759, 2021.

\bibitem{RappaportItWillWork}
T.~S. Rappaport {\em et~al.}, ``Millimeter wave mobile communications for 5g
  cellular: It will work!,'' {\em IEEE Access}, vol.~1, pp.~335--349, 2013.

\bibitem{cassini}
R.~Ferr\'{e}ol, ``{Cassini Oval},'' 2017.
\newblock [Online]. Available:
  \url{https://mathcurve.com/courbes2d.gb/cassini/cassini.shtml}.

\end{thebibliography}
\bibliographystyle{ieeetr}

\vfill

\end{document}